\documentclass[]{aiaa-tc}

\usepackage{bm}
\usepackage{amsmath}
\usepackage[capitalise]{cleveref}
\crefname{equation}{Eq.}{Eqs.}
\crefname{figure}{figure}{figures}
\crefname{table}{table}{tables}
\crefname{section}{section}{sections}
\usepackage{tikz}
\usepackage{amsfonts}
\usepackage{subcaption}
\usepackage[normalem]{ulem}
\tikzstyle{rect} = [rectangle, minimum width=5cm, minimum height=2.5cm, text centered, text width=6cm, draw=black]
\tikzstyle{arrow} = [thick,->,>=stealth]
\tikzstyle{dbl_arrow} = [thick,<->,>=stealth]
\usepackage{fancyhdr}
 \title{Safety-margin-based design and redesign considering mixed epistemic model uncertainty and aleatory parameter uncertainty}

 \author{
  Nathaniel B. Price%
    \thanks{Graduate Research Assistant, Mechanical \& Aerospace Engineering, natprice@ufl.edu}\
  , Mathieu Balesdent%
  \thanks{Research Engineer, mathieu.balesdent@onera.fr}\
  , S\'{e}bastien Defoort%
  \thanks{Research Engineer, sebastien.defoort@onera.fr}
  ,\\ Rodolphe Le Riche%
  \thanks{CNRS Permanent Research Associate, leriche@emse.fr}\
  , Nam H. Kim%
  \thanks{Associate Professor, Mechanical \& Aerospace Engineering, nkim@ufl.edu}\
  , Raphael T. Haftka%
  \thanks{Distinguished Professor, Mechanical \& Aerospace Engineering, haftka@ufl.edu}\
  \\
  {\normalsize\itshape
   \textsuperscript{*\dag\ddag}ONERA - The French Aerospace Lab, Palaiseau, France}\\
  {\normalsize\itshape
   \textsuperscript{*\P\textbardbl}University of Florida, Gainesville, Florida, USA}\\
  {\normalsize\itshape
   \textsuperscript{*\textsection}\'{E}cole des Mines de Saint-\'{E}tienne, Saint-\'{E}tienne, France} \\
  {\normalsize\itshape
   \textsuperscript{\textsection}CNRS LIMOS UMR 6158, Saint-\'{E}tienne, France}
 }

 \AIAApapernumber{YEAR-NUMBER}
 \AIAAconference{Conference Name, Date, and Location}
 \AIAAcopyright{\AIAAcopyrightD{YEAR}}

\tikzstyle{arrow} = [thick,->,>=stealth]
\tikzstyle{dbl_arrow} = [thick,<->,>=stealth]
\tikzstyle{arrow_right} = [thick,|->,>=stealth]
\tikzstyle{line1} = [draw, -latex']
\tikzstyle{line2} = [draw]
\tikzstyle{rect_sm} = [rectangle, minimum width=1in, minimum height=1in, text centered, text width=0.75in, draw=black, fill=blue!10]
\tikzstyle{rect_sm_empty} = [rectangle, minimum width=1in, minimum height=1in, text width=0.75in, fill=white]
\tikzstyle{diam} = [diamond, draw, fill=blue!10, 
text width=2cm, text badly centered, node distance=2cm, inner sep=0pt]

\begin{document}

\maketitle

\begin{abstract}
At the initial design stage engineers often rely on low-fidelity models that have high epistemic uncertainty. Traditional safety-margin-based deterministic design resorts to testing (e.g. prototype experiment, evaluation of high-fidelity simulation, etc.) to reduce epistemic uncertainty and achieve targeted levels of safety. Testing is used to calibrate models and prescribe redesign when tests are not passed. After calibration, reduced epistemic model uncertainty can be leveraged through redesign to restore safety or improve design performance; however, redesign may be associated with substantial costs or delays. In this paper, a methodology is described for optimizing the safety-margin-based design, testing, and redesign process to allow the designer to tradeoff between the risk of future redesign and the possible performance and reliability benefits. The proposed methodology represents the epistemic model uncertainty with a Kriging surrogate and is applicable in a wide range of design problems. The method is illustrated on a cantilever beam bending example and then a sounding rocket example. It is shown that redesign acts as a type of quality control measure to prevent an undesirable initial design from being accepted as the final design. It is found that the optimal design/redesign strategy for maximizing expected design performance includes not only redesign to correct an initial design that is later revealed to be unsafe, but also redesign to improve performance when an initial design is later revealed to be too conservative (e.g. too heavy).
\end{abstract}

\section*{Nomenclature}
\begin{tabbing}
  \hspace*{0.5in} \= \kill
  $\bm{x}$ \> Design variable vector \\
  $\bm{U}$ \> Aleatory random variable vector \\
  $n$ \> Safety margin \\
  $e$ \> Epistemic model error\\
  $f(\cdot,\cdot)$ \> Objective function \\
  $g(\cdot,\cdot)$ \> Limit-state function \\
  $\sigma_G(\cdot,\cdot)$ \> Standard deviation of limit-state function  \\
  $q$ \> Redesign indicator function \\
  $p_{re}$ \> Probability of redesign \\
  $p_{F}$ \> Probability of failure \\
  $\mathbb{E}[\cdot]$ \> Expected value operator \\
  $\mathbb{P}[\cdot]$ \> Probability operator \\
  $\Phi(\cdot)$ \> Normal CDF \\
  $\beta$ \> Reliability index \\
  \\
  \textit{Subscripts}\\
  $L$ \> Low-fidelity model \\
  $H$ \> High-fidelity model \\
  $cons$ \> Conservative deterministic value \\
  $ini$ \> Initial design \\
  $re$ \> Design after redesign \\
  $final$ \> Final design after possible redesign \\
  $lb$ \> Lower bound \\
  $ub$ \> Upper bound \\
  $F$ \> Failure \\
  $U$ \> Aleatory uncertainty \\
  $E$ \> Epistemic uncertainty \\
  \\
  \textit{Superscripts}\\
  $(i)$ \> Epistemic realization \\
  $\star$ \> Target value in optimization \\
  \\
  \textit{Accents}\\
  $\bar{}$ \> Mean value
\end{tabbing}

\section{Introduction}
At the initial design stage engineers usually rely on low-fidelity models that have high epistemic uncertainty. Uncertainty is typically classified as aleatory or epistemic \cite{pate-cornell_uncertainties_1996,ferson_different_1996,oberkampf_error_2002}. Epistemic uncertainty is due to lack of knowledge, is reducible by gaining more information, and has a fixed but unknown value. Aleatory uncertainty is due to variability, is irreducible, and is a distributed quantity. In engineering design, a system is typically designed to be robust with respect to aleatory variables such as enviromental conditions or material variability. The robustness of the system to aleatory uncertainty may be controlled implicitly through safety margins, safety factors\cite{federal_aviation_regulations_25.303_2015}, and conservative design values\cite{federal_aviation_regulations_25.613_2015} or explicitly through reliability-based design methods. However, there are relatively few design methods that consider epistemic model uncertainty \cite{mahadevan_inclusion_2006,jiang_reliability-based_2013,kim_reliability-based_2008}. Errors in low-fidelity models, which may be considered indicative of errors in reliability estimates, are often revealed in the future when higher fidelity simulations are performed or prototypes are tested. If the improved model reveals significant discrepancies between low and high-fidelity simulations or between simulations and prototypes, redesign may be required to correct the initial design.

Redesign, also known as engineering change, is the process of revising an initial design conditional on new knowledge\cite{jarratt_engineering_2011}. Typically, redesign is performed if a low-fidelity model is revealed to have unconservative bias that may indicate an unsafe initial design. Redesign is also beneficial when an initial design is revealed to be overly conservative such that the design performance can be significantly improved. Redesign provides an opportunity for design improvement, however, it is often viewed as a problem in industry because redesign may be associated with substantial costs and delays \cite{wright_review_1997}. Designers could benefit from controlling the probability of future redesign and trading off between the probability of redesign and design performance\cite{villanueva_accounting_2014}. However, predicting how the reliability and performance may change conditional on future redesign is a complex and computationally expensive task.

Even without considering redesign, there is significant computational cost involved in mixed epistemic and aleatory uncertainty propagation. For example in a two level Monte-Carlo simulation (MCS), for each epistemic realization sampled in the outer loop, many aleatory realizations are sampled and propagated through design models in the inner loop in order to calculate a distribution, or family, of distributions \cite{hoffman_propagation_1994}. Two-level uncertainty propagation is computationally costly, but provides the complete distribution of probability of failure which can be used to calculate a variety of useful statistics, such as confidence intervals\cite{pate-cornell_uncertainties_1996}. Alternatively, a model with epistemic model uncertainty could be replaced with a conservative prediction, such as mean plus $k$-standard deviation offset of epistemic model uncertainty, in order to avoid the expensive two-level uncertainty propagation\cite{kim_reliability-based_2008,dubourg_adaptive_2011}. After replacing the uncertain model with the conservative model, the aleatory uncertainty can be propagated through the conservative model as usual. However, the former approach allows for precise reliability statements such as ``we believe with 1-$\alpha$ confidence that the probability of failure is less than $p_F^\alpha$'' whereas the interpretation of the latter approach is less straightforward and may only yield ``pseudo-confidence bounds\cite{dubourg_adaptive_2011}''. The reliability assessment becomes more complex when we consider that the design variables are epistemic random variables. That is, if there is some probability of future redesign then the final design is an epistemic random variable because it is unknown (e.g. incomplete, imprecise, or uncertain specification) at the initial design stage.

In this study, we propose a design method that considers mixed epistemic model uncertainty and aleatory parameter uncertainty and  includes the possibility of future redesign. Both aleatory and epistemic uncertainties are modeled using the probability formalism. It will be shown that redesign acts as a type of quality control measure for epistemic uncertainty by implementing design changes in response to extreme epistemic realizations. In the proposed method, aleatory and epistemic uncertainties in the reliability assessment are handled sequentially rather than in a nested fashion. In a preliminary step, traditional RBDO is performed with respect to aleatory parameter uncertainty using the mean low-fidelity model in order to find the most probable point (MPP) of the aleatory random variables with respect to the mean low-fidelity model. In subsequent steps, aleatory random variables are fixed at this MPP and a $k$ standard deviation offset is used as a safety margin with respect to epistemic model uncertainty. An initial design is found based on deterministic optimization using a standard deviation offset $k_{ini}$. In the future, the initial design will be tested (i.e. the high-fidelity model will be evaluated at the initial design) and the redesign decision will be based on the observed discrepancy between the low and high-fidelity models. If the observed discrepancy is less than $k_{lb}$ or above $k_{ub}$ then redesign will be performed. During redesign a possibly different standard deviation offset $k_{re}$ is used. The outcome of the future high-fidelity evaluation (i.e. future test) is unknown at the initial design stage and therefore the design process is repeated in a MCS. The MCS allows for the calculation of the probability of redesign and a prediction of how future redesign is related to final design performance and reliability. The standard deviation offsets $\bm{k}=\{k_{ini},k_{lb},k_{ub},k_{re}\}$ governing the design process are optimized to minimize the expected value of the objective function while satisfying constraints on reliability and probability of redesign. In contrast to previous work on simulating the effects of a future test and redesign\cite{villanueva_including_2011,villanueva_accounting_2014,matsumura_reliability_2013}, this study accounts for spatial correlations in epistemic model uncertainty by using a Kriging model to represent model uncertainty and significantly reduces the computational cost by proposing a computationally cheap approximation of the reliability constraint. After the optimization of the standard deviation offsets, the complete probability of failure distribution is recovered through two-level uncertainty propagation.

In \cref{sec:methods} the general method of simulating a future test and possible redesign is described. In \cref{sec:test_cases} the method is illustrated on a cantilever beam bending example and then a multidisciplinary sounding rocket design problem. In \cref{sec:conclusion} the study is summarized and the implications of the method and results are discussed.

\section{Methods}
\label{sec:methods}
In this section, the step-by-step procedure of the propose method is explained. In \cref{sec:rbdo}, the conservative values that will be used in place of aleatory random variables are found based on preliminary RBDO. In \cref{sec:rule_opt}, the formulation of the optimization of the standard deviation offsets is presented. The Monte-Carlo simulation (MCS) of epistemic error realizations is described in \cref{sec:sim_multi_fut}. A single sample in the MCS consists of a complete deterministic design / redesign process as described in \cref{sec:det_design}. In \cref{sec:prob_eval}, the calculation of the expected objective function value, probability of redesign, and probability of the probability of failure exceeding a target value are described.

\subsection{Preliminary reliability-based design optimization (RBDO)}
\label{sec:rbdo}
Preliminary reliability-based design optimization (RBDO) is performed using the mean low-fidelity model of the limit-state function and considering only aleatory uncertainty. In subsequent steps, aleatory random variables are fixed at the MPP as the design is optimized deterministically. The preliminary RBDO problem is formulated as
\begin{equation}
\label{eq:rbdo}
\begin{array}{ll}
\min			&  \mathbb{E}_U\left[f(\bm{x},\bm{U})\right] \\
\text{w.r.t}	&  \bm{x} \\
\text{s.t.}		&  \mathbb{P}_U\left[\bar{g}(\bm{x},\bm{U})\le 0\right] \le p_F^{\star} \\
\end{array}
\end{equation} 
where $\mathbb{E}_U[\cdot]$ is an expectation operator with respect to aleatory uncertainty, $\mathbb{P}_U[\cdot]$ is a probability operator with respect to aleatory uncertainty, $f(\cdot,\cdot)$ is the objective function, $\bm{x}\in\mathbb{R}^d$ is a vector of design variables, $\bm{U}$ is a vector of aleatory random variables with a realization $\bm{u}\in\mathbb{R}^p$,  $\bar{g}_{H}(\cdot,\cdot)$ is the mean limit-state function with respect to epistemic model uncertainty, and $p_F^\star$ is the target probability of failure. We assume the limit-state function is formulated such that failure is defined as $g_H(\cdot,\cdot)<0$. The formulation of the search for the MPP of the RBDO optimum $\bm{x}_{RBDO}$ is
\begin{equation}
\label{eq:mpp}
\begin{array}{ll}
\min			&  ||\bm{\hat{u}}|| \\
\text{w.r.t}	&  \bm{\hat{u}} \\
\text{s.t.}		&  \bar{g}(\bm{x}_{RBDO},\bm{\hat{u}}) \ge 0 \\
\end{array}
\end{equation}
where the optimization is performed in standard normal space with $\bm{\hat{u}}$ denoting the transformd variable. Since the RBDO problem does not consider epistemic model uncertainty in the limit-state function there is a high probability that the resulting optimum could be very unsafe or very conservative. However, the computational cost of the optimization problem is much lower than formulating an optimization with full two-level mixed epistemic / aleatory uncertainty propagation. The task of locating a design that is conservative with respect to epistemic model uncertainty, but not overly so, will be addressed in the remainder of the proposed method.

\subsection{Optimization of standard deviation offsets}
\label{sec:rule_opt}
The optimization of the standard deviation offsets (i.e. safety margins) is formulated as
\begin{equation}
\label{eq:saf_opt}
\begin{array}{ll}
\min			& \mathbb{E}_E\left[\mathbb{E}_U\left[f(\bm{X}_{final},\bm{U})\right]\right] \\
\text{w.r.t}	& \bm{k}=\{k_{ini},-k_{lb},k_{ub},k_{re}\} \\
\text{s.t.}		& \mathbb{P}_E\left[P_F(\bm{X}_{final}) \ge p_F^\star\right]\le \alpha \\
				& p_{re} \le p_{re}^\star \\
				& 0.0 \le \bm{k} \le 4.0 \\
\end{array}
\end{equation}
where $\mathbb{E}_E[\cdot]$ an expectation operator with respect to epistemic uncertainty, $\bm{X}_{final}$ is a vector of final optimum design variables after possible redesign, $\mathbb{P}_E[\cdot]$ is a probability operator with respect to epistemic uncertainty, $P_F(\cdot)$ is the probability of failure with respect to aleatory uncertainty, $1-\alpha$ is the desired confidence level, and $p_{re}$ is the probability of redesign. We define the probability of failure for the $i$-th realization of epistemic model uncertainty as $p_F^{(i)}(\bm{x}_{final}^{(i)})=\mathbb{P}_U\left[g^{(i)}(\bm{x}_{final}^{(i)},\bm{U})\le 0\right]$. Note that in the optimization of the standard deviation offsets in \cref{eq:saf_opt} we consider a distribution of probability of failure by calculating a realization of the probability of failure for each realization of epistemic model uncertainty. This is in contrast to the preliminary RBDO problem in \cref{eq:rbdo} where the constraint was defined with respect to the mean model $\bar{g}(\cdot,\cdot)$ and did not consider epistemic uncertainty. The final design, $\bm{X}_{final}$, is uncertain because we consider the possibility that the design may need to be redesigned in the future conditional on the outcome of a high-fidelity evaluation of the initial design. The probability of failure, $P_F(\cdot)$, is uncertain because there is epistemic model uncertainty in the limit-state function and because the design is uncertain. The tradeoff between the expected objective function value and probability of redesign is captured by solving the single objective optimization problem for several values of the constraint $p_{re}^\star$. The global optimization is performed using Covariance Matrix Adaptation Evolution Strategy (CMA-ES) \cite{hansen_cma_2006} with a penalization strategy to handle the constraints.

The computational cost of the standard deviation offsets optimization problem is high due to the mixed epistemic and aleatory uncertainty in the reliability constraint. To reduce the computational cost, the reliability constraint is approximated as
\begin{equation}
\label{eq:pf_approx}
\mathbb{P}_E\left[P_F(\bm{X}_{final}) \ge p_F^\star\right]\approx \mathbb{P}_E\left[G_H(\bm{X}_{final},\bm{u}_{cons})\le 0\right]
\end{equation}
where $\bm{u}_{cons}$ is a vector of fixed conservative values used in place of aleatory variables corresponding to the MPP as found in \cref{eq:rbdo,eq:mpp} and $G_H(\cdot,\cdot)$ is an uncertain limit-state function due to epistemic model error. The true probability on the left-hand side of \cref{eq:pf_approx} requires two-level uncertainty propagation, but the approximation on the right only considers epistemic uncertainty and is therefore only requires single level uncertainty propagation. The approximation is inspired by studies on reliability-based design considering only aleatory uncertainty where the reliability constraint is converted to an equivalent deterministic constraint \cite{wu_efficient_1998, wu_safety-factor_2001, du_sequential_2004} There are two elements that contribute the the error in the proposed approximation. First, the MPP is an epistemic random variable due to model uncertainty so any single point estimate will incur some degree of error. Second, the final design is an epistemic random variable and will differ from $\bm{x}_{RBDO}$ where the MPP search was performed. It is assumed that the MPP with respect to the mean limit-state function $\bar{g}(\cdot,\cdot)$ is a reasonable approximation of the mean MPP with respect to the realizations of the uncertain limit-state function $G(\cdot,\cdot)$. That is, it is assumed the MPP of the mean is close to the mean of the MPP's. Furthermore, it is assumed that the distribution of final designs $\bm{X}_{final}$ will be centered near $\bm{x}_{RBDO}$. These approximations are reasonable when the nonlinearity of the limit-state function is not significant. The approximation is introduced to reduced the cost of the optimization of the standard deviation offsets. The full two-level uncertainty propagation is performed after the convergence of \cref{eq:saf_opt} for the optimum standard deviation offsets in order to recover the full probability of failure distribution and assess the accuracy of the approximation. If the full two-level uncertainty propagation reveals significant error in the approximation of the reliability constraint for the optimum design, then this approximation should not be used in the optimization of the standard deviation offsets. For the two test cases in this study, the approximation is shown to yield reasonable results.

\subsection{Monte-Carlo simulation of epistemic model error}
\label{sec:sim_multi_fut}
The epistemic model uncertainty and aleatory parameter uncertainty are treated separately (see \cite{hoffman_propagation_1994, pate-cornell_uncertainties_1996, helton_treatment_1994}). The true relationship between the different fidelity models is assumed to be of the form
\begin{equation}
\label{eq:true}
g_{H}(\bm{x},\bm{u})=g_{L}(\bm{x},\bm{u})+e(\bm{x},\bm{u})
\end{equation}
where $g_{H}(\cdot,\cdot)$ is the high-fidelity model, $g_{L}(\cdot,\cdot)$ is the low-fidelty model, and $e(\cdot,\cdot)$ is the error between the low-fidelity and high-fidelity models. Typically, the error $e(\cdot,\cdot)$ is unknown. For example, in the cantilever beam bending example presented later in this study the low-fidelity model is based on Euler-Bernoulli beam theory and the high-fidelity model is based on Timoshenko beam theory and the error or discrepancy between the two beam models is assumed to be unknown. The uncertainty in the model error is represented as a Kriging model $E(\cdot,\cdot)$. Based on the possible model errors the high-fidelity model is predicted as 
\begin{equation}
\label{eq:true_pred}
G_{H}(\bm{x},\bm{u})=g_{L}(\bm{x},\bm{u})+E(\bm{x},\bm{u})
\end{equation}
The Kriging model for the error is constructed in the joint space of the aleatory variables, $\bm{u}$, and the design variables, $\bm{x}$. The uncertainty in $G_{H}(\bm{x},\bm{u})$ in \cref{eq:true_pred} is only due to epistemic model error $E(\cdot,\cdot)$. Propagation of aleatory uncertainty $\bm{U}$ through the uncertain model is discussed in \cref{sec:prob_eval}. For simplicity of notation, we will define the mean of the Kriging prediction for the error as $\bar{e}(\cdot,\cdot)$ and the standard deviation as $\sigma_E(\cdot,\cdot)$. The mean prediction of the high-fidelity model is
\begin{equation}
\bar{g}_{H}(\bm{x},\bm{u})=g_{L}(\bm{x},\bm{u})+\bar{e}(\bm{x},\bm{u})
\end{equation}
with standard deviation $\sigma_G(\cdot,\cdot)=\sigma_E(\cdot,\cdot)$.

The epistemic random function $E(\cdot,\cdot)$ is used to represent the lack of knowledge regarding how well the low-fidelity model matches the high-fidelity model. Assuming initial test data is available, maximum likelihood estimation (MLE) will be used to estimate the parameters of the Kriging model. The prediction $G_{H}(\cdot,\cdot)$ is viewed as a distribution of possible functions. Samples or trajectories drawn from this distribution that are conditional on initial test data are referred to as conditional simulations. In the absence of test data these realizations are unconditional simulations. Let $g_{H}^{(i)}(\cdot,\cdot)$ denote the i-th realization of $G_{H}(\cdot,\cdot)$ based on a realization $e^{(i)}(\cdot,\cdot)$ of the Kriging model $E(\cdot,\cdot)$. A variety of methods exist for generating these conditional simulations \cite{chiles_chapter_2012}. In this study, the conditional simulations are generated directly based on Cholesky factorization of the covariance matrix using the STK Matlab toolbox for Kriging \cite{bect_stk:_2014} and by sequential conditioning \cite{chiles_chapter_2012}. 

We can consider a Monte-Carlo simulation of $m$ conditional simulations $i=1,\dots,m$ corresponding to $m$ possible futures. In practice, the sample size $m$ is increased until the estimated coefficient of variation of the quantity of interest is below a certain threshold. The design process conditional on one error realization is described in \cref{sec:det_design}. By repeating the design process for many different error realizations (i.e. for different possible high-fidelity models through \cref{eq:true_pred}) we can determine the distribution of possible final design outcomes.

\subsection{Deterministic Design Process}
\label{sec:det_design}
The deterministic design process is controlled by a vector of standard deviation offsets $\bm{k}$. The design process consists of finding an initial design, testing the initial design by evaluating it with the high-fidelity model, and possible calibration and redesign. The future high-fidelity evaluation of the initial design (i.e. future test) is unknown and therefore modeled as an epistemic random variable. The redesign decision, calibration, and redesign optimum are conditional on a particular test result. In \cref{sec:ini_design,sec:testing,sec:redesign} the process is described conditional on the $i$-th error realization $E(\cdot,\cdot)=e^{(i)}(\cdot,\cdot)$. 

\subsubsection{Initial design}
\label{sec:ini_design}
The design problem is formulated as a deterministic optimization problem
\begin{equation}
\label{eq:ddo}
\begin{array}{ll}
\min			&  f(\bm{x},\bm{u}_{cons}) \\
\text{w.r.t}	&  \bm{x} \\
\text{s.t.}		&  \bar{g}_H(\bm{x},\bm{u}_{cons})- k_{ini}\sigma_G(\bm{x},\bm{u}_{cons}) \ge 0 \\
\end{array}
\end{equation}
where $\bar{g}_{H}(\cdot,\cdot)$ is the mean of the predicted high-fidelity model, $k_{ini}$ is the initial standard deviation offset, $\bm{u}_{cons}$ is a vector of conservative deterministic values used in place of aleatory random variables, and $\sigma_G(\cdot,\cdot)$ is the standard deviation of the limit-state function with respect to epistemic model uncertainty. Let $\bm{x}_{ini}$ denote the optimum design found from \cref{eq:ddo}. There is no uncertainty in the initial design $\bm{x}_{ini}$ because the optimization problem is defined using the mean of the model prediction and fixed conservative values, $\bm{u}_{cons}$, are used in place of aleatory random variables.

\subsubsection{Testing initial design and redesign decision}
\label{sec:testing}
A possible high-fidelity evaluation, $g_{H}^{(i)}(\bm{x}_{ini},\bm{u}_{cons})$, of the initial design $\bm{x}_{ini}$ is simulated. The test will be passed if $n_{lb}\le g_{H}^{(i)}(\bm{x}_{ini},\bm{u}_{cons}) \le n_{ub}$ where $n_{lb}$ and $n_{ub}$ correspond to lower and upper bounds on acceptable safety margins. The redesign decision can be formulated in terms of standard deviation offsets as $k_{lb}\le z_{ini}^{(i)} \le k_{ub}$ where 
\begin{equation}
Z_{ini}=\frac{G_H(\bm{x}_{ini},\bm{u}_{cons})-\bar{g}_H(\bm{x}_{ini},\bm{u}_{cons})}{\sigma_G(\bm{x}_{ini},\bm{u}_{cons})}
\end{equation}
If the observed safety margin is too low ($g_{H}^{(i)}(\bm{x}_{ini},\bm{u}_{cons})<n_{lb}$) then the design is unsafe and redesign should be performed to restore safety. If the observed safety margin is too high ($g_{H}^{(i)}(\bm{x}_{ini},\bm{u}_{cons})>n_{ub}$) then the design is too conservative and it may be worth redesigning to improve performance. Let $q^{(i)}$ denote an indicator function for the redesign decision that is 1 for redesign and 0 otherwise. We will refer to redesign triggered by a low safety margin (i.e. $n_{lb}$) as redesign for safety and redesign triggered by a high safety  margin (i.e. $n_{ub}$) as redesign for performance. If the test is not passed then redesign should be performed to select a new design.

\subsubsection{Calibration and redesign}
\label{sec:redesign}
If redesign is required, the model is first calibrated conditional on the test result. To obtain the calibrated model, the test realization $g_{H}^{(i)}(\bm{x}_{ini},\bm{u}_{cons})$ corresponding to the error instance $e^{(i)}(\bm{x}_{ini},\bm{u}_{cons})$ is treated as a new data point and the error instance is added to the design of experiment for the error model. The redesign problem is formulated as a deterministic optimization problem
\begin{equation}
\label{eq:ddo_re}
\begin{array}{ll}
\min			&  f(\bm{x},\bm{u}_{cons}) \\
\text{w.r.t}	&  \bm{x} \\
\text{s.t.}		&  \bar{g}_{H,calib}^{(i)}(\bm{x},\bm{u}_{cons})- k_{re}\sigma_{G,calib}^{(i)}(\bm{x},\bm{u}_{cons}) \ge 0 \\
\end{array}
\end{equation}
where the mean of the predicted high-fidelity model $\bar{g}_{H,calib}^{(i)}(\cdot,\cdot)$ and the standard deviation $\sigma_{G,calib}^{(i)}(\cdot,\cdot)$ are calibrated conditional on the test result $g_{H}^{(i)}(\bm{x}_{ini},\bm{u}_{cons})$ and $k_{re}$ is a new standard deviation offset. Let $\bm{x}_{re}^{(i)}$ denote the optimum design after redesign found from \cref{eq:ddo_re}. Comparing the initial design problem in \cref{eq:ddo} to the redesign problem in \cref{eq:ddo_re}, we see that there is a change in the feasible design space due to the change in the standard deviation offset and calibration. Note that the calibration is conditional on obtaining the high-fidelity evaluation $g_{H}^{(i)}(\bm{x}_{ini},\bm{u}_{cons})$ in the future. That is, if we obtain the evaluation $g_{H}^{(i)}(\bm{x}_{ini},\bm{u}_{cons})$, we can obtain the calibrated model $\bar{g}_{H,calib}^{(i)}(\cdot,\cdot)$, and we will select an improved design $\bm{x}_{re}^{(i)}$.

\subsection{Probabilistic Evaluation}
\label{sec:prob_eval}
The final design after possible redesign is
\begin{equation}
\bm{x}_{final}^{(i)}=\left(1-q^{(i)}\right)\bm{x}_{ini}+q^{(i)}\bm{x}_{re}^{(i)}
\end{equation}
where $q^{(i)}=1$ corresponds to failing the test and performing redesign. The expected objective function value after possible redesign is $\mathbb{E}_E\left[\mathbb{E}_U\left[f(\bm{X}_{final},\bm{U})\right]\right]$. 
The probability of redesign is calculated analytically as 
\begin{equation}
p_{re}=\Phi(k_{lb})+(1-\Phi(k_{ub}))
\end{equation}
where $\Phi(\cdot)$ is the standard normal cumulative distribution function (cdf).  

The optimization of the standard deviation offsets is based on a computationally cheap approximation of the reliability constraint as described in \cref{sec:rule_opt}. The key benefit of the proposed approximation is that the probability can be calculated analytically. The probability of a negative safety margin conditional on passing the test and keeping the initial design is
\begin{equation}
\mathbb{P}_E\left[G(\bm{x}_{ini},\bm{u}_{cons}) \le 0 | Q=0 \right] =\Phi_{T}(-k_{ini})
\end{equation}
where $\Phi_{T}(\cdot)$ is the normal cdf truncated to the interval $[-k_{lb},k_{ub}]$. The probability conditional on performing redesign is
\begin{equation}
\mathbb{P}_E\left[G(\bm{X}_{re},\bm{u}_{cons}) \le 0 | Q=1\right] =\Phi(-k_{re})
\end{equation}
The final probability of a negative  safety margin after possible redesign is
\begin{equation}
\label{eq:pf_n_exact}
\mathbb{P}_E\left[G(\bm{X}_{final},\bm{u}_{cons}) \le 0 \right]= (1-p_{re})\Phi_{T}(-k_{ini})+p_{re}\Phi(-k_{re})
\end{equation}

After solving the optimization problem in \cref{eq:saf_opt}, the full two-level mixed aleatory / epistemic uncertainty propagation is performed to recover the probability of failure distribution and check the accuracy of the proposed approximation. The probability of failure of the final design is unknown since there is epistemic uncertainty in the model $G_H(\cdot,\cdot)$. A realization of the probability of failure is calculated conditional on an error realization $E(\cdot,\cdot)=e^{(i)}(\cdot,\cdot)$. A realization of the probability of failure of the initial design is
\begin{equation}
\label{eq:pf_ini}
p_F^{(i)}(\bm{x}_{ini})=\mathbb{P}_U\left[g_{H}^{(i)}(\bm{x}_{ini},\bm{U}) < 0\right]
\end{equation}
where $\mathbb{P}_U\left[\cdot\right]$ denotes the probability with respect to aleatory uncertainty. Note that the epistemic model uncertainty is treated separately from the aleatory uncertainty to distinguish between the quantity of interest, the probability of failure with respect to the high-fidelity model and aleatory uncertainty, and the lack of knowledge regarding this quantity. The error in the low-fidelity model $E(\cdot,\cdot)$ has no impact on the reliability with respect to the high-fidelity model $g_{H}(\cdot,\cdot)$. However, since the high-fidelity model is unknown, the probability of failure calculation is repeated many times conditional on many different realizations of the high-fidelity model $g_{H}^{(i)}(\cdot,\cdot)$ through \cref{eq:pf_ini}. A realization of the final probability of failure after possible redesign is
\begin{equation}
\label{eq:pf_re}
p_F^{(i)}(\bm{x}_{re}^{(i)})=\mathbb{P}_U\left[g_{H}^{(i)}(\bm{x}_{re}^{(i)},\bm{U})\le 0\right]
\end{equation}
After redesign, the design variable $\bm{x}_{re}^{(i)}$ is also an epistemic random variable in addition to the limit state function $g_{H}^{(i)}(\cdot,\cdot)$. Many different methods are available for calculating the probability of failure. In this study, first order reliability method (FORM) is used to calculate the probability of failure for each epistemic realization. The final probability of failure after possible redesign is 
\begin{equation}
p_F^{(i)}(\bm{x}_{final}^{(i)})=\left(1-q^{(i)}\right)p_F^{(i)}(\bm{x}_{ini})+q^{(i)}p_F^{(i)}(\bm{x}_{re}^{(i)})
\end{equation}
Note that the redesign decision $q^{(i)}$ shapes the final probability of failure distribution because we will have the opportunity in the future to correct the initial design if it fails the deterministic test. The probability of the probability of failure of the final design exceeding the target probability of failure is estimated by MCS as
\begin{equation}
\label{eq:pf_mcs}
\mathbb{P}_E\left[P_F(\bm{X}_{final}) \ge p_F^\star\right]\approx \frac{1}{m}\sum_{i=1}^{m}{I\left[p_F^{(i)}(\bm{x}_{final}^{(i)}) \ge p_F^\star\right]}
\end{equation}
where $I[\cdot]$ is an indicator function. The computational cost of the full two-level mixed aleatory / epistemic uncertainty propagation is high and therefore only performed after the optimization of the standard deviation offsets. For example, more than $m=1900$ probability of failure calculations are necessary to estimate a probability of the order $\alpha=0.05$ with a 10\% coefficient of variation.

\section{Test cases}
\label{sec:test_cases}
\subsection{Cantilever beam bending example}
\label{sec:beam}
\subsubsection{Problem description}
The first example is the design of a cantilever beam to minimize mass subject to a constraint on tip displacement adapted from an example by Wu et al\cite{wu_safety-factor_2001}. The beam is subject to independent aleatory random loads in the horizontal and vertical directions. The original problem involved the design of a long slender beam and therefore used Euler-Bernoulli beam theory. In this example, the length of the beam is reduced such that shear stress effects become important and Timoshenko beam theory is more accurate. The Timoshenko beam model plays the role of a computationally expensive high-fidelity model (e.g. finite element analysis) and the Euler-Bernoulli beam model plays the role of an inexpensive low-fidelity model. The beam is optimized to ensure with 95\% confidence that the reliability index of the final design after possible redesign is greater than 3.

The low-fidelity model of the limit state function is
\begin{equation}
g_L(\bm{x},\bm{U})=d^\star-\frac{4l^3}{ewt}\sqrt{\left(\frac{F_Y}{t^2}\right)^2+\left(\frac{F_X}{w^2}\right)^2}
\end{equation}
where $\bm{x}=\{w,t\}$ are the design variables and $\bm{U}=\{F_X,F_Y\}$ are the aleatory variables. The high-fidelity model of the limit state function is 
\begin{equation}
g_H(\bm{x},\bm{U})=d^\star-\sqrt{(d_x(\bm{x},\bm{U}))^2+(d_y(\bm{x},\bm{U}))^2}
\end{equation}
where $d_x$ and $d_y$ are given by \cref{eq:dx,eq:dy}. The problem parameters are described in \cref{tab:exParam}.
\begin{equation}
\label{eq:dx}
d_x(\bm{x},\bm{U})=\left(\frac{3lF_X}{2gwt}+\frac{4l^3F_X}{ewt^3}\right)
\end{equation}
\begin{equation}
\label{eq:dy}
d_y(\bm{x},\bm{U})=\left(\frac{3lF_Y}{2gwt}+\frac{4l^3F_Y}{ew^3t}\right)
\end{equation}
The objective function is the cross-sectional area of the beam
\begin{equation}
f(\bm{x})=wt
\end{equation}
which is proportional to the mass of the beam.

\begin{table*}
\centering
\caption{\label{tab:exParam} Parameters for cantilever beam example}
\begin{tabular}{llll}
 						     & Parameter                  & Notation & Value \\\hline
Design variables, $\bm{x}$   & Width of cross section     & $w$      & $2.5\le w \le 5.5$ in   \\
						     & Thickness of cross section & $t$      & $1.5\le t \le 4.5$ in\\
Aleatory variables, $\bm{U}$ & Horizontal load            & $F_X$    & $N(500,100^2)$ lbs   \\
							 & Vertical load              & $F_Y$    & $N(1000,100^2)$ lbs   \\
Constants					 & Elastic modulus            & $e$      & $29\times 10^6$ psi \\
							 & Shear modulus           	  & $g$      & $11.2\times 10^6$ psi \\
							 & Length of beam			  & $l$		 & $10$ in \\
							 & Allowable tip displacement & $d^\star$& $2.25 \times 10^{-3}$ in \\
							 & Conservative aleatory values & $\bm{u}_{cons}$ & $\{744.7, 1173.5\}$ lbs \\
							 & Target probability of failure & $p_F^\star=\Phi(-\beta^\star)$ & $1.35\times 10^{-3}=\Phi(-3)$ \\
							 & Target confidence level		 & $1-\alpha$ & $0.95$ \\
\end{tabular}
\end{table*}

\begin{figure*}
\centering
\includegraphics[width=5in]{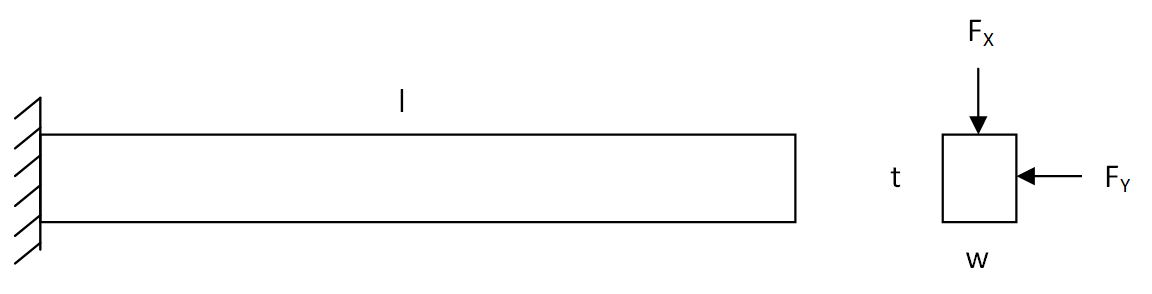}
\caption{The beam is subject to horizontal and vertical tip loads}
\label{fig:beam_example}
\end{figure*}

\subsubsection{Application of the proposed method}
\begin{figure*}
\centering
\begin{subfigure}{3.25in}
  \centering
  \includegraphics[width=1\textwidth]{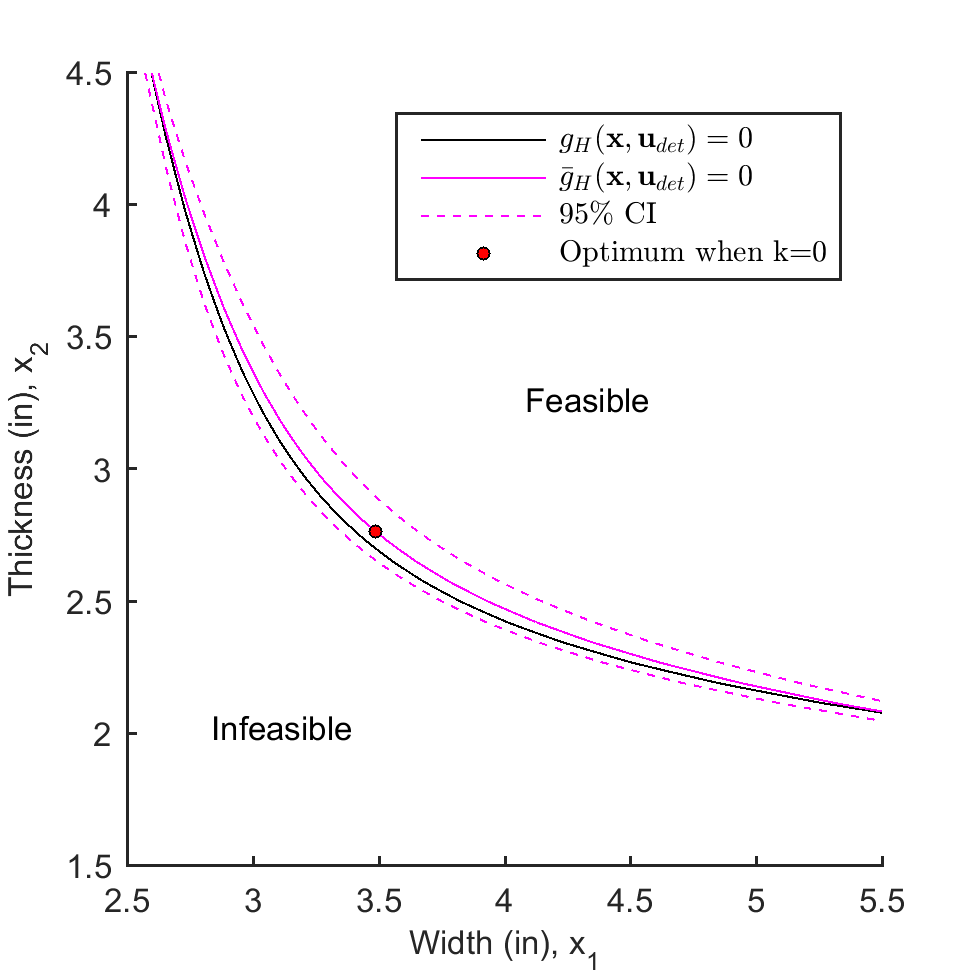}
  \caption{Design space}
  \label{fig:design_opt}
\end{subfigure}%
\begin{subfigure}{3.25in}
  \centering
  \includegraphics[height=1\textwidth]{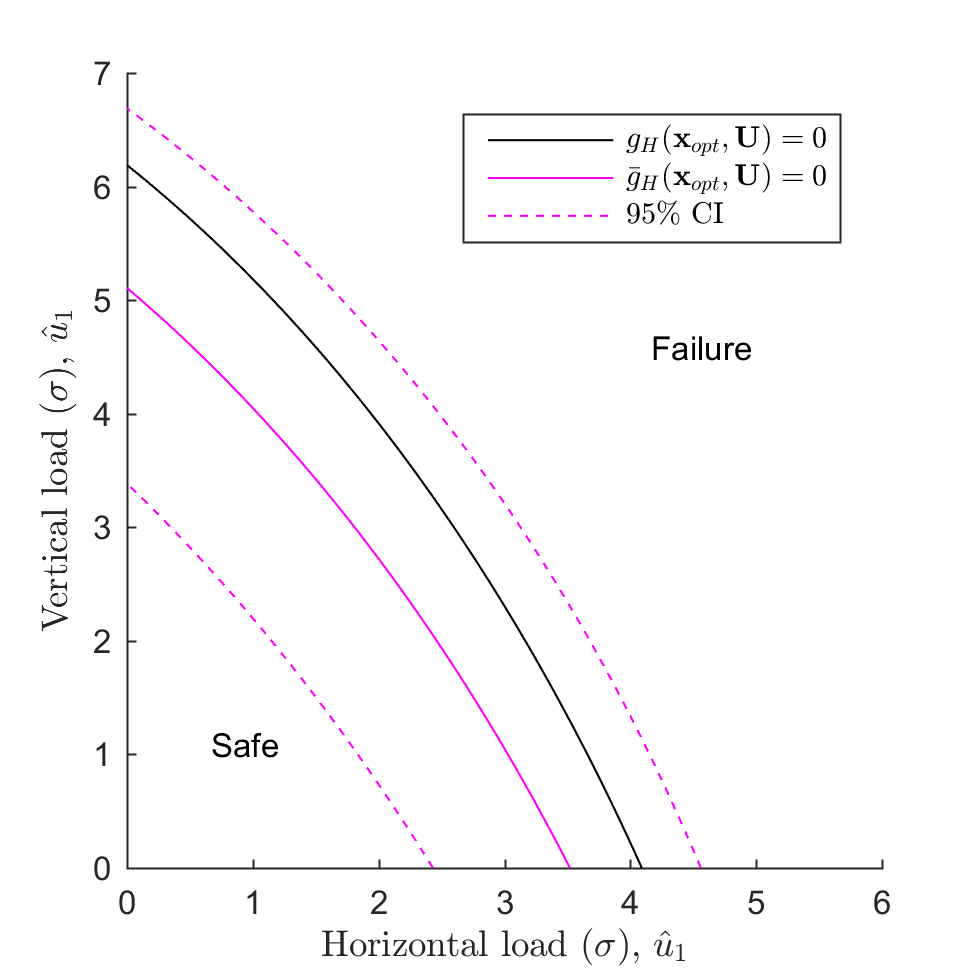}
  \caption{Standard normal aleatory space}
  \label{fig:reliability_analysis}
\end{subfigure}
\caption{The figure on the left shows the design optimization with standard deviation offset $k=0$ and fixed conservative values $\bm{u}_{cons}=\{744.7, 1173.5\}$ lbs in place of aleatory variables. The figure on the right shows the limit-state function in standard normal space for the optimum design found on the left.}
\label{fig:design_opt_reliability_analysis}
\end{figure*}

\paragraph{Step 1: Quantifying the model uncertainty}
The first step is to quantify the uncertainty in the low-fidelity model. A Kriging model is constructed for the discrepancy between the low and high-fidelity models based on evaluations at the corner points in the joint design-aleatory space (4 beam designs each with 4 loading conditions). To demonstrate the method, the corner points were chosen in order to ensure high model uncertainty. In practice, the model could also be constructed based on data from previous designs. The Kriging model improves the prediction from the low-fidelity model, but more importantly it provides confidence intervals for the model uncertainty. In \cref{fig:design_opt_reliability_analysis}, the confidence intervals arising due to model uncertainty are shown in the design space and aleatory space.

\paragraph{Step 2: Selecting fixed conservative values for aleatory variables}
Next, aleatory random variables $\bm{U}$ are replaced with fixed conservative values $\bm{u}_{cons}$. The conservative values are found by solving the RBDO problem problem in \cref{eq:rbdo}. The RBDO is performed with respect to aleatory uncertainty conditional on the mean low-fidelity model. By solving the optimization problem in \cref{eq:rbdo}, we select conservative values $\bm{u}_{cons}=\{744.7, 1173.5\}$ lbs. These values correspond to approximately the 99th and 96th percentiles of the loads. The RBDO problem only requires single level uncertainty propagation since epistemic model uncertainty is fixed at the mean prediction.

\begin{figure*}
\centering
\includegraphics[height=3.25in]{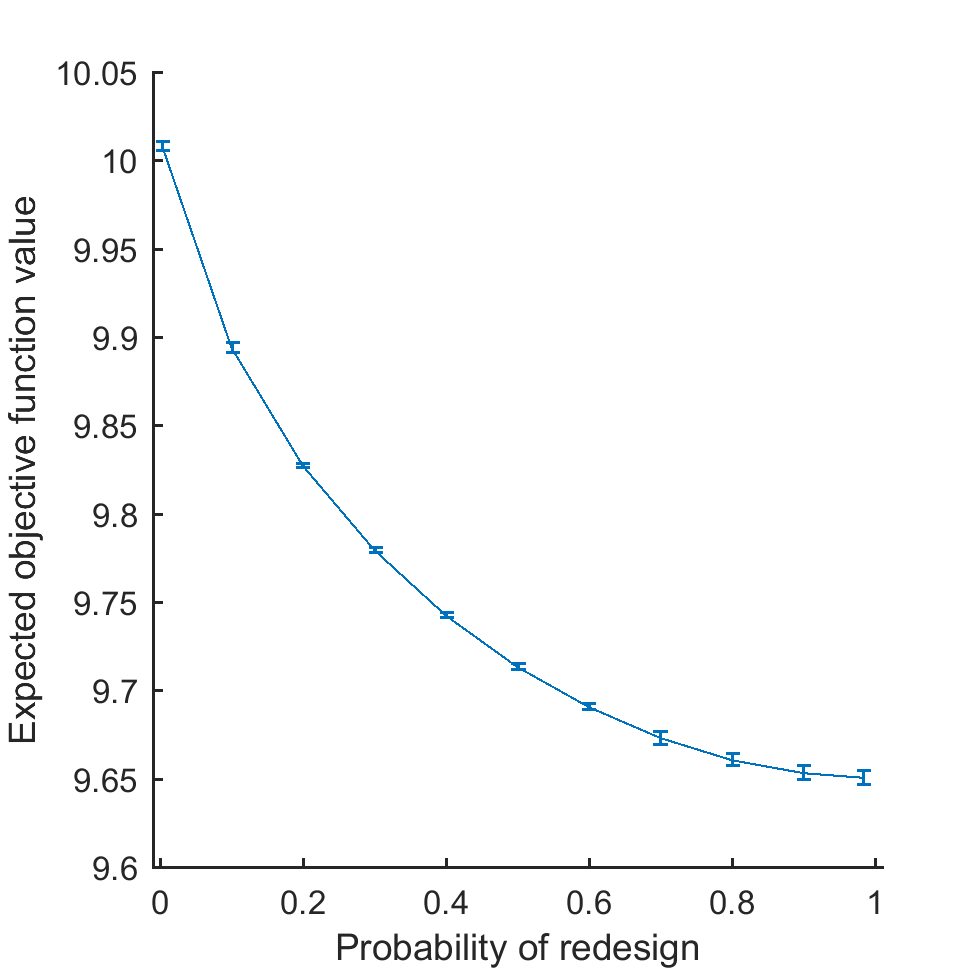}
\caption{Tradeoff curve for expected cross sectional area versus probability of redesign}
\label{fig:beam_tradeoff_curves}
\end{figure*}

\paragraph{Step 3: Optimization of safety margins (i.e. standard deviation offsets)}
In the third step, the optimum standard deviation offsets are found by solving \cref{eq:saf_opt} using CMA-ES with a penalized objective function. The optimization is performed using CMA-ES because the problem is noisy and the standard deviation offsets are optimized globally. Recall that standard deviation offsets of model uncertainty are used during the design / redesign process as safety margins against model uncertainty. Inside the MCS, the design optimization (\cref{eq:ddo,eq:ddo_re}) is performed using sequential quadratic programming (SQP). By varying the constraint on the probability of redesign $p_{re}^\star$ we obtain a curve for the expected cross sectional area versus probability of redesign as shown in \cref{fig:beam_tradeoff_curves}. The tradeoff curve is used to determine how much risk of redesign is acceptable given the expected performance improvement. For illustration, we will select the optimum safety margins $\bm{k}=\{0.71,0.89,2.25,3.00\}$ corresponding to 20\% probability of redesign for more detailed study.

\paragraph{Step 4: Full two-level mixed uncertainty propagation}
In the fourth step, the full two-level mixed uncertainty propagation is performed for the selected optimum safety margins. The full two-level mixed uncertainty propagation is used to recover the probability of failure distribution and obtain detailed results for the MCS of the design/redesign process. In the previous step involving the optimization of the safety margins, aleatory variables were fixed and only epistemic model uncertainty was considered. In the full two-level mixed uncertainty propagation, the probability of failure is calculated using first order reliability method (FORM) for each realization of epistemic model uncertainty (i.e. Kriging conditional simulation) 

\paragraph{Step 5: Post-processing of simulation results}
Finally, post-processing is performed for the data gathered in the MCS. 

\begin{figure*}
\centering
\begin{subfigure}{3.25in}
  \centering
  \includegraphics[width=1\textwidth]{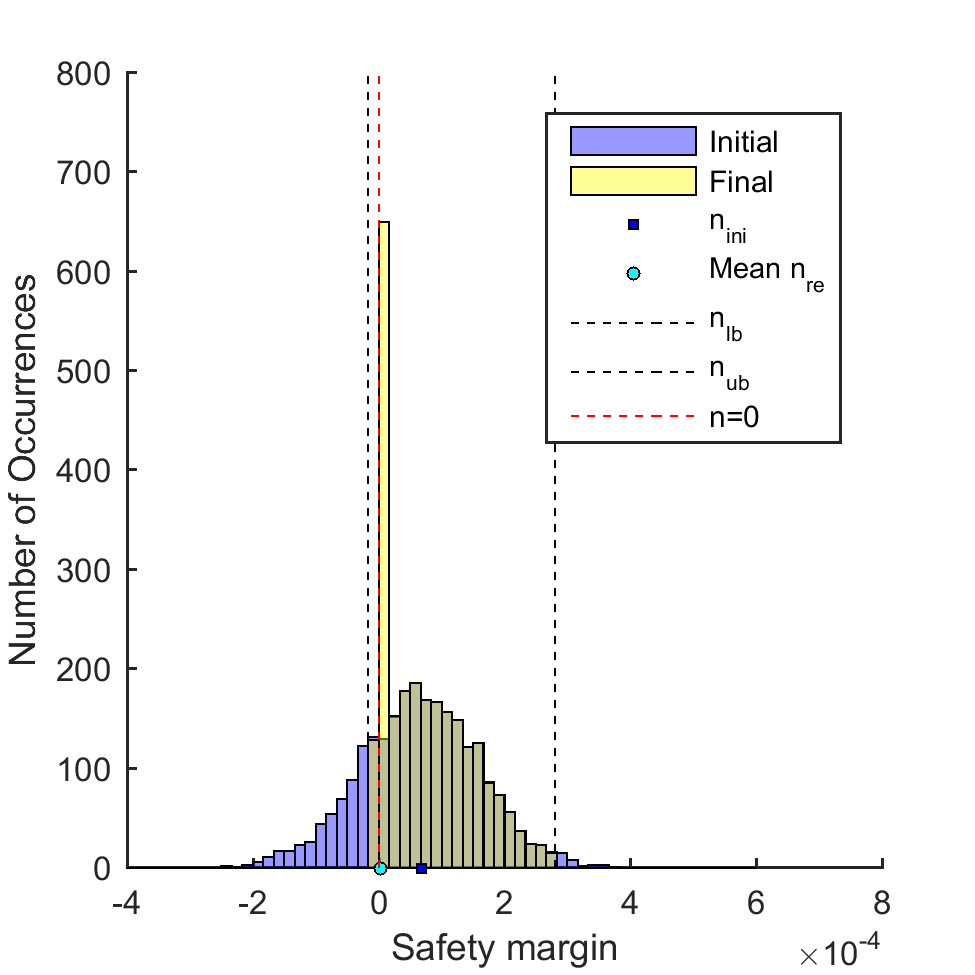}
  \caption{Safety margin}
  \label{fig:beam_margin}
\end{subfigure}%
\begin{subfigure}{3.25in}
  \centering
  \includegraphics[width=1\textwidth]{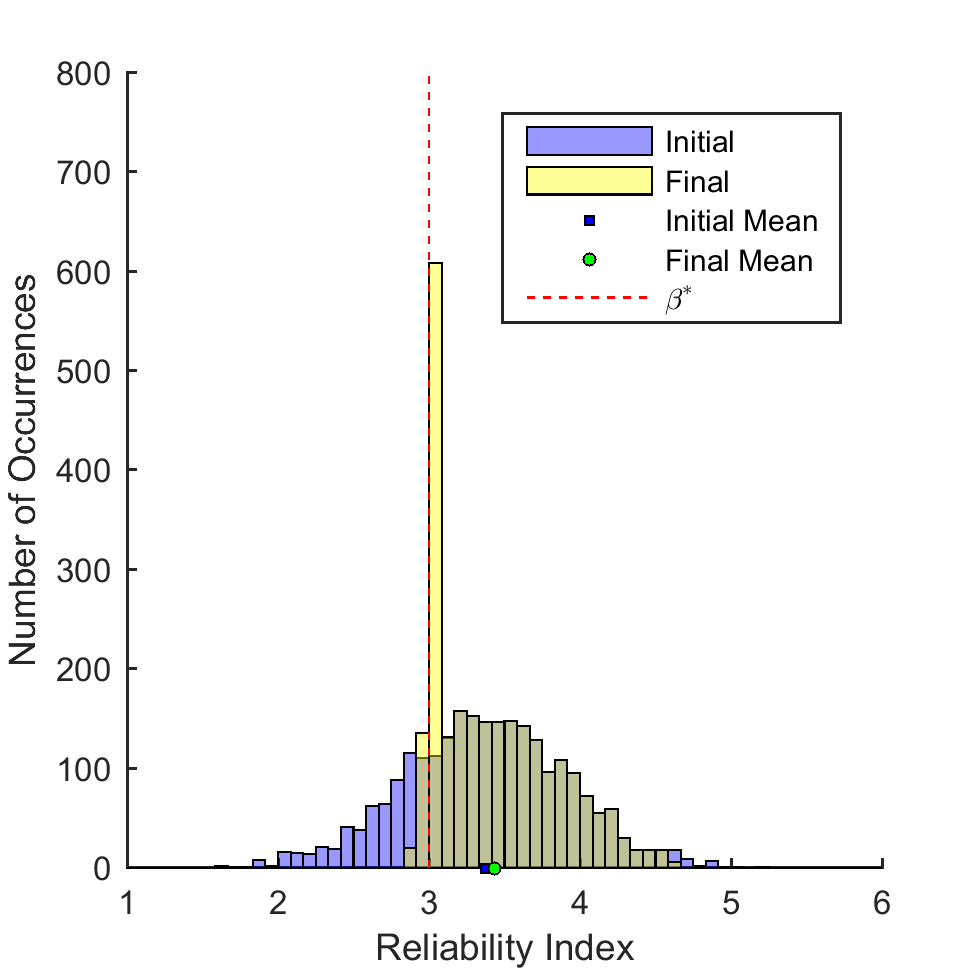}
  \caption{Reliability index}
  \label{fig:beam_beta}
\end{subfigure}
\caption{Distribution of safety margin and reliability index for 20\% probability of redesign. Plots show overlapping transparent histograms.}
\label{fig:beam_margin_beta}
\end{figure*}

\begin{figure*}
\centering
\begin{subfigure}{1.0in}
  \centering
  \includegraphics[width=1\textwidth]{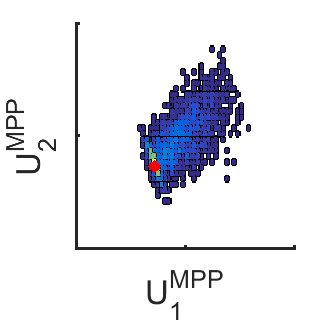}
\end{subfigure}
\begin{subfigure}{1.0in}
  \centering
  \includegraphics[width=1\textwidth]{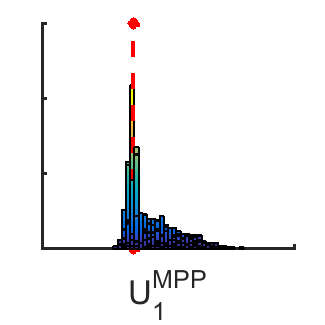}
\end{subfigure}
\begin{subfigure}{1.0in}
  \centering
  \includegraphics[width=1\textwidth]{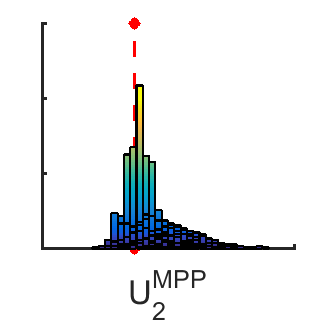}
\end{subfigure}

\centering
\begin{subfigure}{3.25in}
  \centering
  \includegraphics[width=1\textwidth]{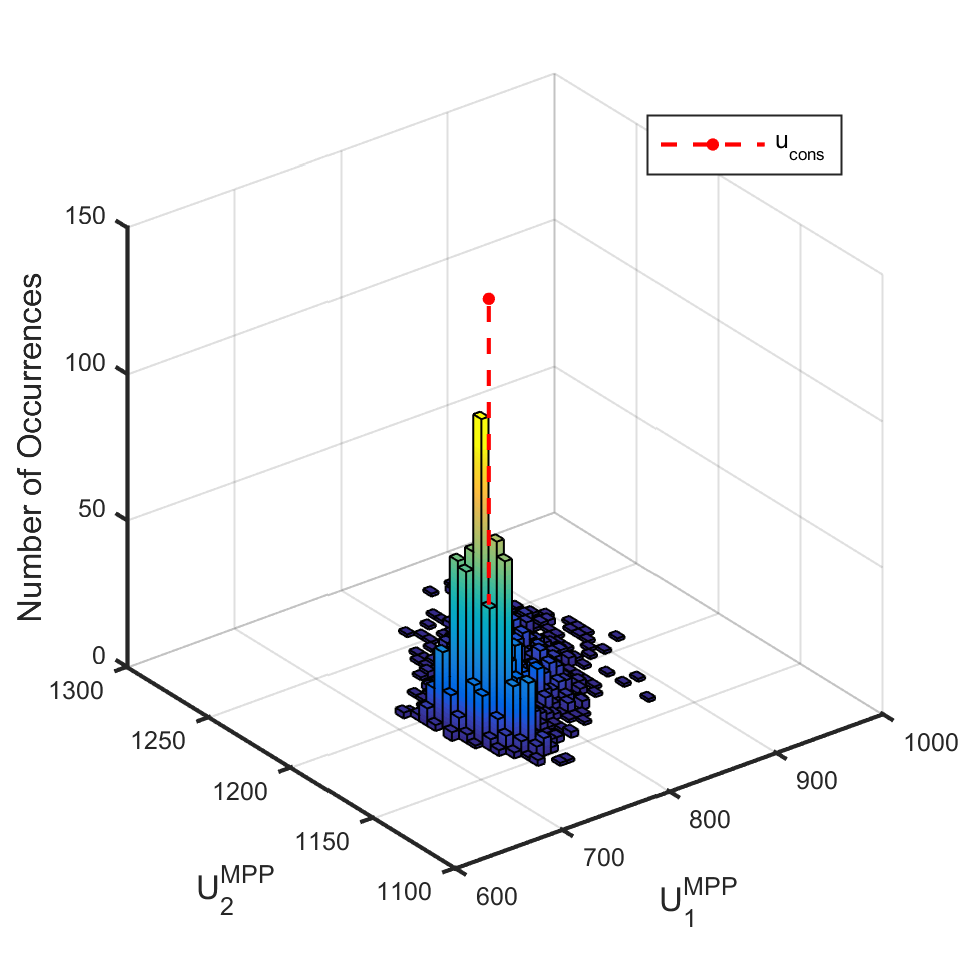}
\end{subfigure}%
\caption{Distribution of most probable point (MPP) for 20\% probability of redesign.}
\label{fig:beam_mpp}
\end{figure*}

First, we examine the safety margin distribution and the reliability index distribution shown in \cref{fig:beam_margin_beta}. The safety margin distribution in \cref{fig:beam_margin} shows the possible constraint violations with respect to epistemic model uncertainty conditional on the fixed conservative values $\bm{u}_{cons}$. The beam will be redesigned if the safety margin is less than $-0.16\times 10^{-4}$ inches or greater than $2.8\times 10^{-4}$ inches. Note that a negative safety margin is possible because the margin is calculated using the conservative aleatory values $\bm{u}_{cons}$. In this example, a negative safety margin indicates the tip displacement is greater than the allowable under the conservative loading $\bm{u}_{cons}$. It can be observed that if redesign is required, we expect to have much more precise control over the tip displacement of the beam due to the knowledge gained from the future test. Redesign acts as a type of quality control measure by initiating design changes in response to observing an extreme safety margin. We can compare the safety margin distribution in \cref{fig:beam_margin} to the reliability index distribution in \cref{fig:beam_beta}. There is a strong correlation between the observed safety margin and the reliability index (correlation coefficient 0.999). As a result, the safety margin based redesign criteria is very useful for identifying overly conservative or unsafe designs. The safety margin is strongly correlated with the reliability index because the safety margin is calculated with respect to the MPP of the mean low-fidelity model. As shown in \cref{fig:beam_mpp}, the conservative values $\bm{u}_{cons}$ provide a reasonable point estimate of the MPP distribution. The standard deviation offsets have been optimized based on the computationally cheap approximation of the reliability constraint in \cref{eq:pf_approx} such that the probability of a negative safety margin after possible redesign is 5\%. After performing the full two-level mixed uncertainty propagation, the probability of the probability of failure exceeding the target value of $1.35\times 10^{-3}$ is estimated to be in the range of 5\% to 7\% (95\% confidence interval with $m=2500$). In other words, we have between 93\% and 95\% confidence that the probability of failure of the final design after possible redesign will be less than $p_F^\star=1.35\times 10^{-3}$.

\begin{figure*}
\begin{subfigure}{3.25in}
  \centering
  \includegraphics[width=1\textwidth]{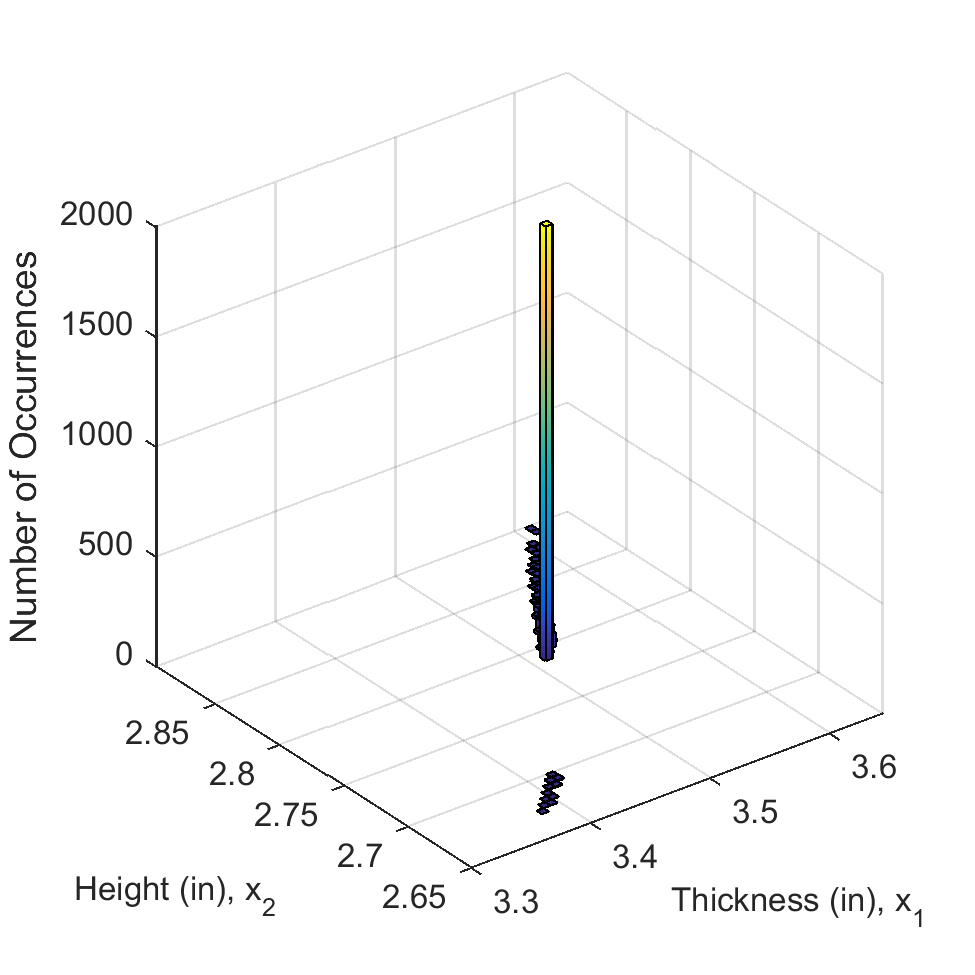}
  \caption{Optimum design variables}
  \label{fig:beam_design}
\end{subfigure}%
\begin{subfigure}{3.25in}
  \centering
  \includegraphics[width=1\textwidth]{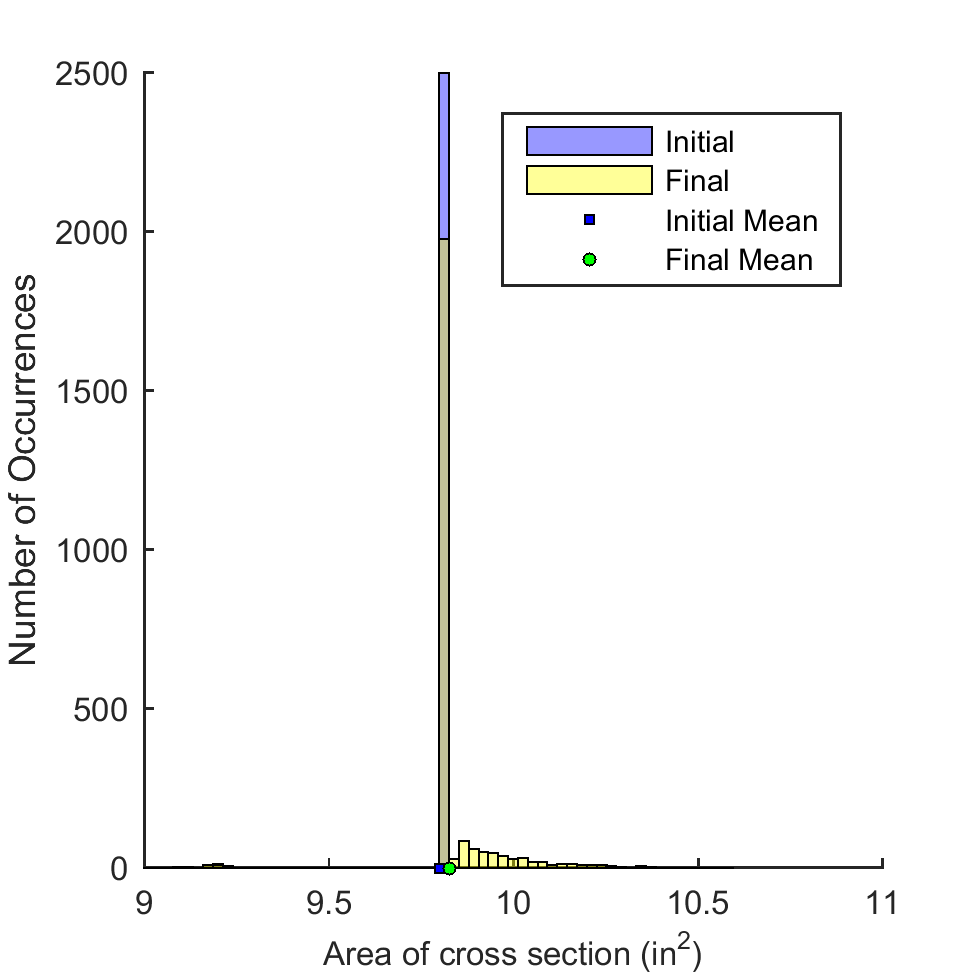}
  \caption{Design performance}
  \label{fig:beam_perf}
\end{subfigure}
\caption{Distribution of optimum design variables and design performance for 20\% probability of redesign. Peak is located at initial design.}
\label{fig:beam_design_perf}
\end{figure*}

Second, we examine the optimum design variable distribution and the cross sectional area distribution shown in \cref{fig:beam_design_perf}. The design variable distribution in \cref{fig:beam_design} shows how the design variables will change if redesign is required in the future. The peak corresponds to the initial design since there is an 80\% probability the initial design will be accepted as the final design. The distribution of design variables can be used to plan for future design changes. The cross sectional area distribution corresponding to the designs in \cref{fig:beam_design} is shown in \cref{fig:beam_perf}. Although the change in the mean area due to all possible design changes is relatively small, the realizations of the area corresponding to redesign may be significantly different than the initial area. For example, if redesign for performance is required the area is reduced by about 6.4\%, however, there is only about a 1\% chance of redesign for performance. On the other hand, there is about a 19\% chance of redesign for safety which is associated with an increase in area of approximately 2\%.


\subsection{Multidisciplinary sounding rocket design example}
\label{sec:rocket}
\subsubsection{Problem description}
\begin{figure}
\centering
\begin{tikzpicture}[x=1in,y=1in]

\draw [dashed,-] (0.625,0.625) -- (4.875,0.625) -- (4.875,-3.875) -- (0.625,-3.875) -- (0.625,0.625);

\node (inputs) [rect_sm_empty, yshift=0in, xshift=0in] {
\small{\textcolor{blue}{Design Variables:}}\\
\scriptsize{\textcolor{blue}{$M_p$: propellant mass}}\\
\scriptsize{\textcolor{blue}{$T/W$: thrust to weight ratio}} \\
\scriptsize{\textcolor{blue}{$p_{cc}$: chamber pressure}} \\
\scriptsize{\textcolor{blue}{$\alpha_P$: mixture ratio}} \\
\scriptsize{\textcolor{blue}{$D$: diameter}} \\
\scriptsize{\textcolor{blue}{$\eta$: engine efficiency factor}} \\
};

\node (coup) [rect_sm_empty,minimum width=2in,text width=2in, yshift=-3in, xshift=1.25in] {
\small{\textcolor{black}{Coupling Variables:}}\\
\scriptsize{$T$: thrust} \\
\scriptsize{$I_{sp}$: specific impulse} \\
\scriptsize{$q$: mass flow rate} \\
\scriptsize{$M_{inert}$: inert mass} \\
\scriptsize{$C_D$: drag coefficient} \\
\scriptsize{$M$: Mach number} \\
\scriptsize{$n_{ax}^{max}$: maximum axial acceleration} \\
\scriptsize{$P_{dyn}^{max}$: maximum dynamic pressure} \\
\scriptsize{$\delta$: inert mass fraction} \\
\scriptsize{$A_t$: throat area} \\
\scriptsize{$A_e$: exhaust area} \\
};

\node (outputs) [rect_sm_empty, yshift=-3in, xshift=5.5in] {
\small{\textcolor{red}{Outputs:}}\\
\scriptsize{\textcolor{red}{$M_{tot}$: total mass}} \\
\scriptsize{\textcolor{red}{$r_{final}$: maximum altitude}} \\
\scriptsize{\textcolor{red}{$L/D$: length to diameter ratio}} \\
};

\node (prop) [rect_sm, yshift=0in, xshift=1.25in] {
\footnotesize{Propulsion}};
\node (struct) [rect_sm, yshift=-1in, xshift=2.25in] {
\footnotesize{Structures \& Sizing}};
\node (aero) [rect_sm, yshift=-2in, xshift=3.25in] {
\footnotesize{Aerodynamics}};
\node (traj) [rect_sm, yshift=-3in, xshift=4.25in] {
\footnotesize{Trajectory}};

\draw [->] (prop) -| (traj) node [pos=0.5, above] (TextNode) {\footnotesize{$T$, $I_{sp}$, $q$, $A_e$}};
\draw [->] (prop) -| (struct) node [pos=0.5, above] (TextNode) {\footnotesize{$T$,$p_{cc}$,$A_t$}};
\draw [->] (prop) -| (aero) node [pos=0.5, above] (TextNode) {\footnotesize{$T$}};
\draw [->] (struct) -| (traj) node [pos=0.5, above left] (TextNode) {\footnotesize{$M_{tot}$}};
\draw [->] (struct) -| (prop) node [pos=0.5, left] (TextNode) {\footnotesize{$\delta$}};
\draw [->] (aero) -| (traj) node [pos=0.5, above left] (TextNode) {\footnotesize{$C_{D}$}};
\draw [->] (aero) -| (struct) node [pos=0.5, below] (TextNode) {\footnotesize{$n_{ax}^{max}$, $P_{dyn}^{max}$}};
\draw [->] (traj) -| (aero) node [pos=0.5, below] (TextNode) {\footnotesize{$M$}};

\draw [red,->] (traj) -- (outputs);
\draw [blue,->] (inputs) -- (prop);

\end{tikzpicture}
    \caption{Design structure matrix for sounding rocket design example. There are couplings between propulsion/structures, aerodynamics/structures, and trajectory/aerodynamics.}
    \label{fig:dsm_orig}
\end{figure}
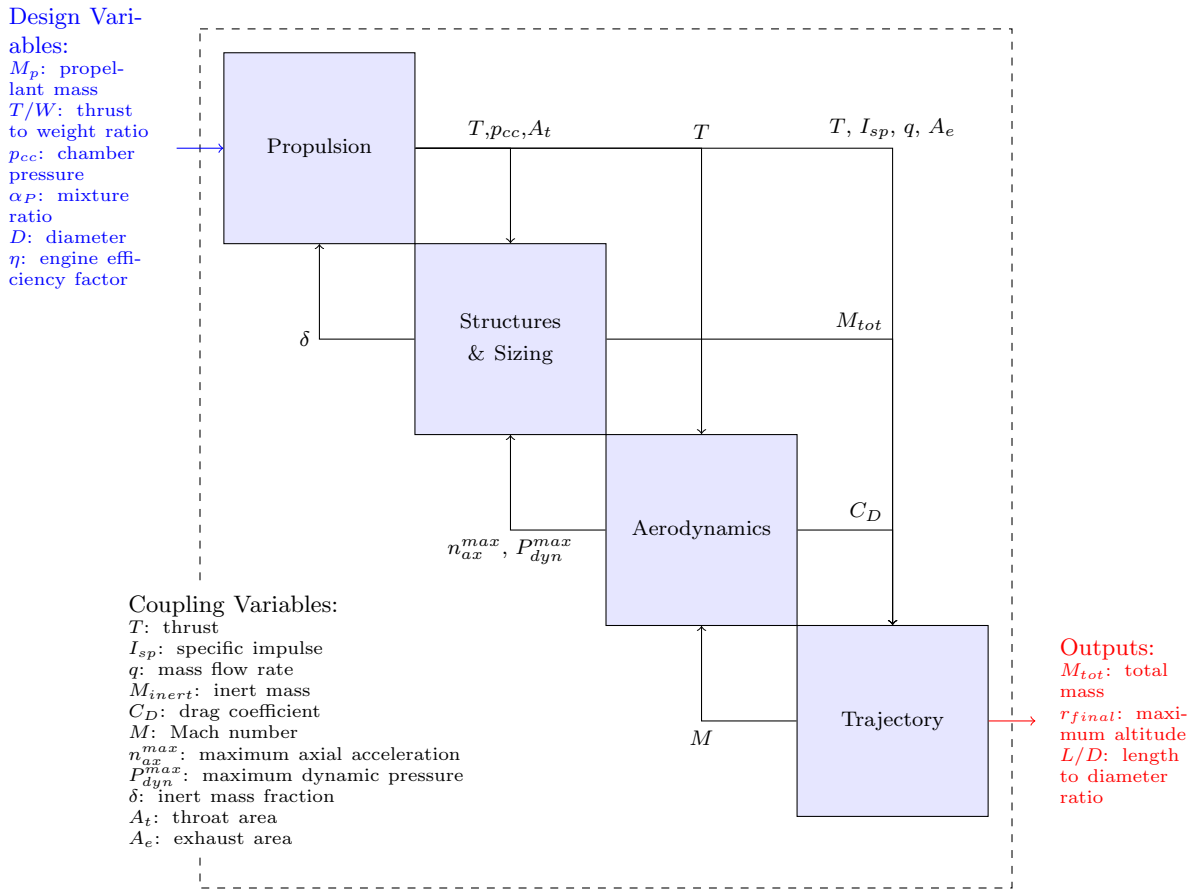

The sounding rocket design example is based on a multidisciplinary design optimization (MDO) problem. The sounding rocket has a single cryogenic liquid hydrogen fueled gas generator engine. The intertank and thrust frame are made from a composite material. The thrust vector control (TVC) system is electromechanical. The avionics and electrical power system have no redundancies. The rocket is designed for vertical integration. The design structure matrix for the sounding rocket example is shown in \cref{fig:dsm_orig}. The analysis uses NASA standard atmosphere models \cite{nasa_us_1976}. There are four disciplines corresponding to propulsion, structures (sizing and weights estimation), aerodynamics, and trajectory simulation. There are five design variables corresponding to the mass of propellant $M_P$, initial thrust to weight ratio $T/W$, engine chamber pressure $p_{cc}$, mixture ratio $\alpha_P$, and diameter $D$. The engine efficiency factor $\eta$ is considered to be an aleatory random variable. The outputs are the total mass $M_{tot}$, final altitude at the end of the propulsion phase $r_{final}$, and length to diameter ratio $L/D$. The design problem is to minimize the total mass while satisfying constraints on the final altitude and the length to diameter ratio. The constraint on the length to diameter ratio is purely deterministic and is therefore simply included as an additional design constraint in the design optimization problems in \cref{eq:ddo,eq:ddo_re}. There is aleatory uncertainty in the final altitude and total mass (GLOW) due to the aleatory uncertainty in the engine efficiency factor $\eta$.

There is a coupling between the structures and aerodynamic disciplines in that the maximum axial acceleration and maximum dynamic pressure are related to the total mass. The structure must be sized to withstand the loads, but changes in the total mass are related to the loads through trajectory and aerodynamics. There is a coupling between structures and propulsion in that the inert mass fraction is related to the thrust through the thrust to weight ratio. The engine mass and thrust frame mass must be designed for a given thrust, but because the thrust to weight ratio is specified beforehand changes in mass alter the thrust. A fixed point iteration is performed to satisfy the coupling constraints with respect to the maximum axial load, maximum dynamic pressure, and inert mass fraction. There is a loop between aerodynamics and trajectory because the drag coefficient varies with Mach number.

\subsubsection{Discipline models}
\label{sec:discipline_models}
The discipline models are mainly based on the dissertation of Castellini, ``Multidisciplinary design optimization for expendable launch vehicles'' \cite{castellini_multidisciplinary_2012}. Full details of the models can be found in the dissertation. The discipline models are briefly summarized here.

\paragraph{Propulsion}
The propulsion discipline calculates the performance characteristics of the engine based on NASA computer program CEA (Chemical Equilibrium with Applications) for calculating chemical equilibrium compositions and properties of complex mixtures \cite{gordon_computer_1994, mcbride_computer_1996}. In order to reduce computational cost, Kriging surrogate models were fit to the characteristic velocity ($C_*$) and thrust coefficient ($C_T$) as a function of mixture ratio, chamber pressure, and nozzle expansion ratio. The surrogate models were constructed based on a design of experiment consisting of 500 points generated using Latin-hypercube sampling. The Kriging models used a Gaussian covariance function and zero order trend functions. Kriging models were constructed in Matlab using DACE (Design and Analysis of Computer Experiments) Matlab toolbox \cite{lophaven_dace_2002}. Any epistemic model uncertainty introduced by the Kriging surrogates in the propulsion discipline is not included in the analysis. The design of experiment size has been set to reduce the model error to quasi zero and the surrogate model is considered as perfect. The specific impulse is calculated as
\begin{equation}
I_{sp}=\frac{C_*C_T \eta}{g_0}
\end{equation}
where $C_*$ is the Kriging prediction of the characteristic velocity, $C_T$ is the the Kriging prediction of the thrust coefficient, $\eta$ is an efficiency factor, and $g_0$ is the standard acceleration due to gravity. The single efficiency factor represents the combined degrading effects of chamber and nozzle losses as well as mass flow losses. The throat area is calculated as
\begin{equation}
A_t=\frac{T}{C_T p_{cc}}
\end{equation}
where $T$ is the thrust and $p_{cc}$ is the chamber pressure. The exhaust area is calculated as
\begin{equation}
A_e=\varepsilon A_t
\end{equation}
where $\varepsilon$ is the nozzle expansion ratio. The mass flow rate is calculated as
\begin{equation}
q=\frac{T}{C_*C_T}=\frac{T}{I_{sp}g_0}
\end{equation}

\paragraph{Structures \& Sizing}
The structures and sizing discipline calculates the total inert mass of the rocket and the total length of the rocket. For this example, the structures and sizing discipline is defined as the combination of sizing and weights estimation. The weights estimation includes engine mass, thrust frame mass, tank mass including thermal protection system, thrust vector control (TVC), and avionics and electrical power system. The thrust frame and tanks are designed using structural safety margins of $SSM=1.1$. All weight estimation relationships (WER's) are based on the dissertation of Castellini and are analytical empirical formulas that are calibrated with existing launch vehicle stages \cite{castellini_multidisciplinary_2012}. The total mass of the rocket is calculated as
\begin{equation}
M_{tot}=M_{inert}+M_P+M_{PL}
\end{equation}
where $M_{inert}$ is the total inert mass, $M_P$ is the propellant mass, and $M_{PL}$ is the payload mass. 

\begin{figure}
\centering
\includegraphics[width=0.5\textwidth]{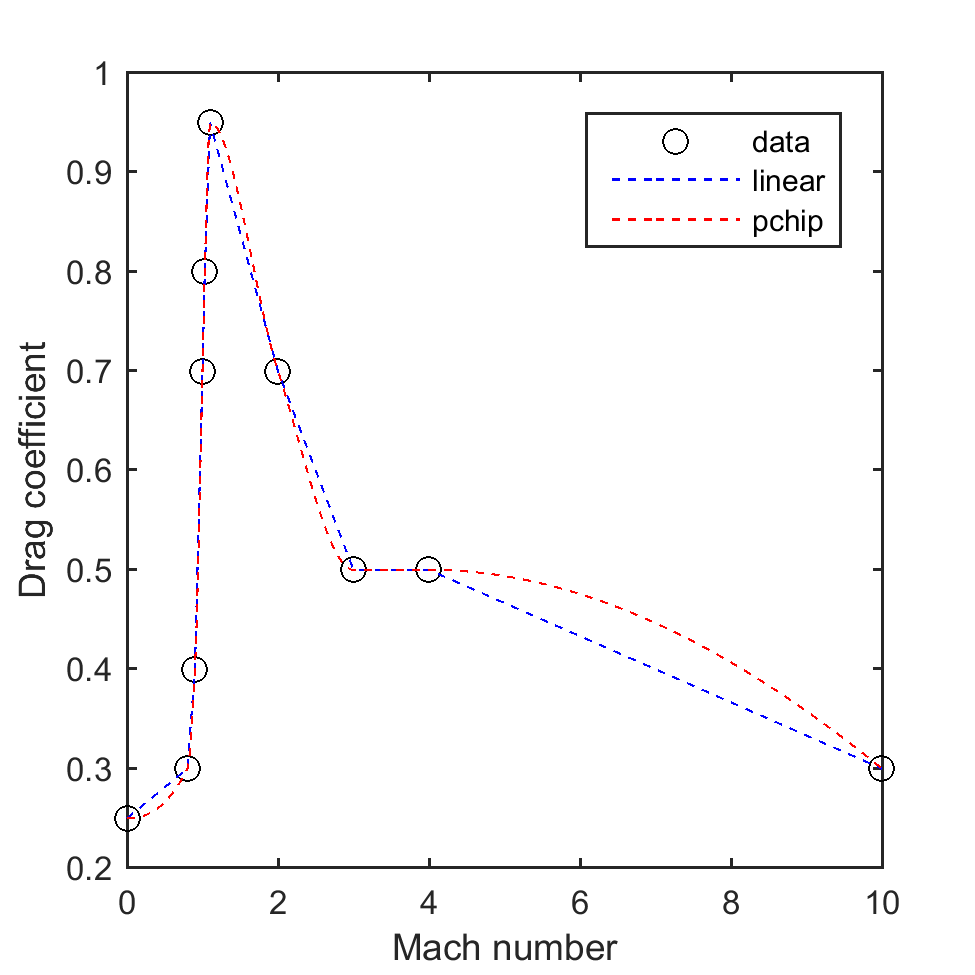}
\caption{Drag coefficient as a function of Mach number based on Missile DATCOM. PCHIP interpolation is used between data points.}
\label{fig:aero_model}
\end{figure}

\paragraph{Aerodynamics}
Given the instantaneous velocity, altitude, and total mass of the rocket the aerodynamics discipline calculates the drag force, dynamic pressure, and axial acceleration. The aerodynamics discipline analysis is based on Missile DATCOM \cite{blake_missile_1998}. In order to reduce computational cost, the drag coefficient is calculated as a function of the Mach number based on PCHIP (piecewise cubic hermite interpolating polynomial) interpolation between values in a table of Missile DATCOM evaluations. The interpolation between data points for the drag coefficient as a function of Mach number is shown in \cref{fig:aero_model}.

The Mach number is calculated as
\begin{equation}
M=\frac{V}{c(r)}
\end{equation}
where the speed of sound $c(r)$ varies as a function of altitude. The axial accelerations in g's is calculated as
\begin{equation}
n_{ax}=\frac{1}{mg_0}\left(T-F_D\right)
\end{equation}
where $F_D=0.5\rho(r) V^2 C_D A$ is the drag force and the air density $\rho(r)$ decreases with altitude. The thrust is calculated as
\begin{equation}
T=I_{sp}g_0 q - A_e P_a(r)
\end{equation}
where $A_e$ is the exhaust area and the air pressure $P_a(r)$ decreases with altitude. The dynamic pressure is calculated as
\begin{equation}
P_{dyn}=0.5\rho(r) V^2
\end{equation}

\paragraph{Trajectory}
The trajectory discipline calculates the altitude, velocity, and total mass as a function of time. The trajectory discipline analysis is based on a two dimensional model. The equations of motion are
\begin{equation}
\begin{array}{l}
\dot{r}=V \\
\dot{V}=\frac{1}{m}\left(-F_D + T - \frac{G M_E m}{r^2}\right) \\
\dot{m}=-q
\end{array}
\end{equation}
where $r$ is the radius, $V$ is the norm of the velocity vector, $F_D$ is the drag force, $T$ is the thrust, $G$ is the gravitational constant, $M_E$ is the mass of the earth, and $m$ is the mass of the rocket. Equations of motion are derived assuming the flight path angle ($\gamma$) and pitch angle ($\theta$) are both 90 degrees. The trajectory discipline is coupled with the aerodynamics discipline. During integration, the trajectory discipline calls the aerodynamics discipline to update the instantaneous values of the thrust and drag force.

\subsubsection{Low-fidelity model}
\begin{figure}
\centering
\includegraphics[width=0.5\textwidth]{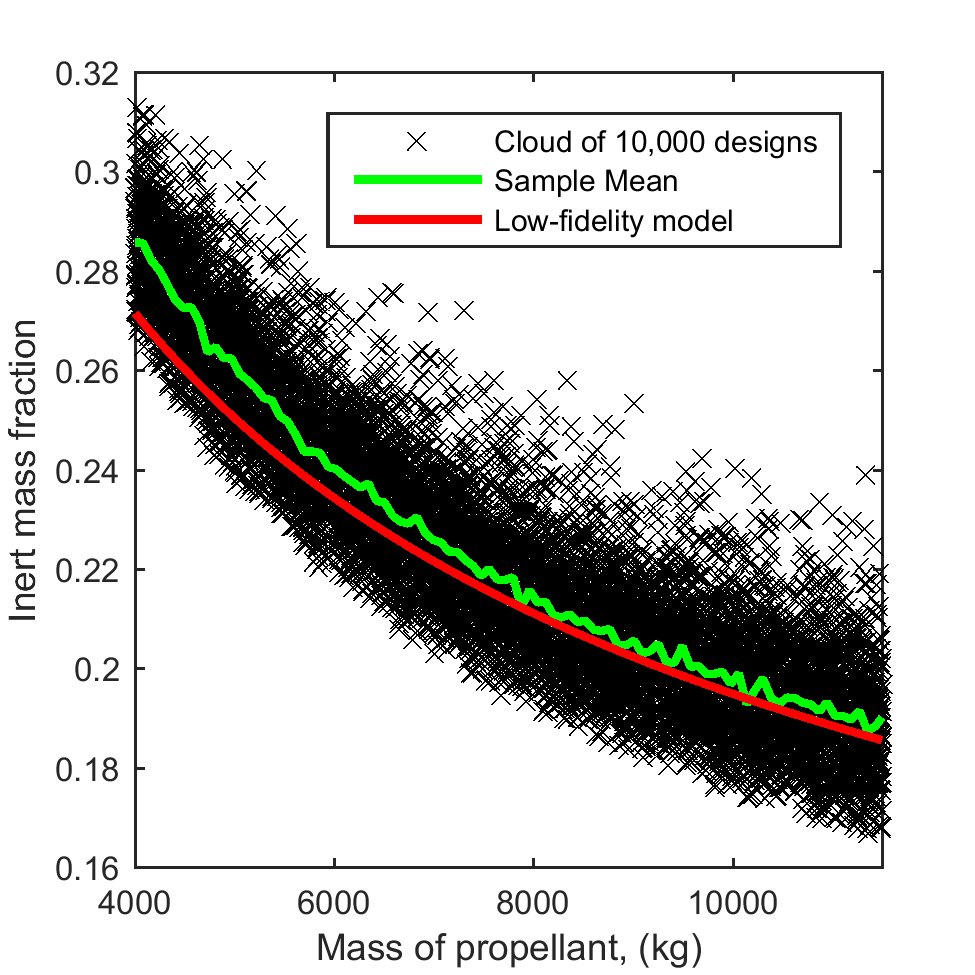}
\caption{A cloud of 10,000 designs in 6-dimensions is projected onto a one dimensional plane and compared to the low-fidelity model prediction}
\label{fig:lf_model_cloud}
\end{figure}

A low-fidelity approximation is introduced for the inert mass fraction as a function of the mass of propellant. The low-fidelity model is based on a curve fit of the model provided in the ``Handbook of Cost Engineering and Design of Space Transportation'' \cite{koelle_handbook_2013}. Table \ref{tab:lf_data} lists the data that was read from the figure (approximated visually). A second order polynomial was fit to the inert mass fraction as a function of the log of propellant
\begin{equation}
\label{eq:delta_L}
\delta_L=(1.5879\log(M_P)^2-36.1554 \log(M_P) + 217.8084)/100
\end{equation}
The design curve is for rockets that are much larger than the sounding rocket we are investigating in this design example. Therefore, we will extrapolate outside of the range of the design curve using the polynomial curve fit. The extrapolation may introduce significant error on top of the already questionable accuracy of the low-fidelity model. The low-fidelity mass model is a 1-dimensional function. However, in the fully coupled system the mass depends on all 6 design-aleatory variables. To visualize the accuracy of the low-fidelity model, a cloud of 10,000 different designs was generated in the 6-dimensional design-aleatory space using Latin-hypercube sampling. Fixed point iterations were performed for each of the designs to enforce coupling constraints between disciplines. In \cref{fig:lf_model_cloud}, the 10,000 designs are projected onto a 1-dimensional plane in order to compare with the 1-dimensional low-fidelity model. It is observed that the low-fidelity model captures the overall trend, but there is significant error. Furthermore, there appears to be significant scatter in the design points around the mean trend line. This is because different designs are being projected onto the 1-dimensional plane. The low-fidelity model is incapable of representing this variation with respect to design variables other than the mass of propellant.

\begin{table}
\centering
\caption{Data read from design curve}
\label{tab:lf_data}
\begin{tabular}{ll}
Mass of propellant (kg) & Inert Mass Fraction \\\hline
10,000 & 0.195 \\
20,000 & 0.155 \\
30,000 & 0.138 \\
40,000 & 0.130 \\
50,000 & 0.125 \\
\end{tabular}
\end{table}

\begin{figure}
\centering
\includegraphics[width=0.5\textwidth]{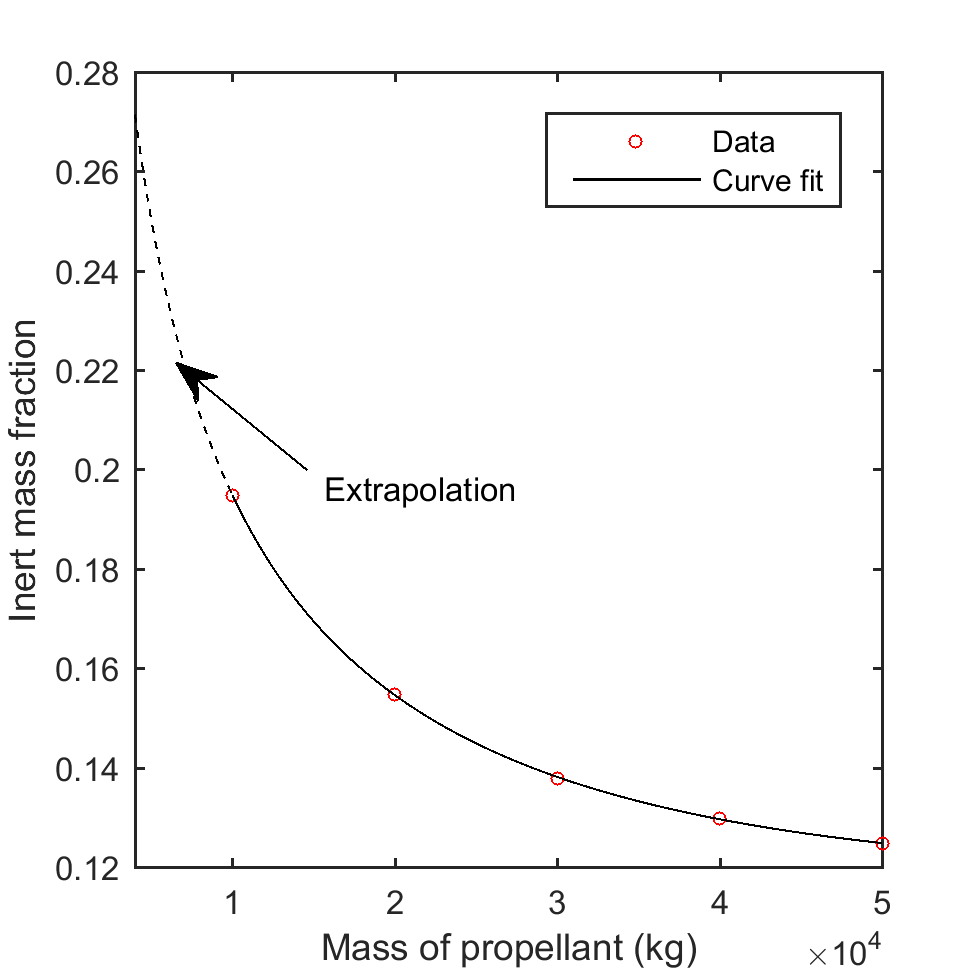}
\caption{A second order polynomial was fit to the inert mass fraction as a function of the log of the propellant mass. The model is extrapolated to the region of interest for sounding rocket design.}
\label{fig:lf_model}
\end{figure}

\subsubsection{Application of the proposed method}
\paragraph{Step 1: Quantifying the model uncertainty}
The first step is to quantify the uncertainty in the low-fidelity model. The low-fidelity model of the inert mass fraction is related to the high-fidelity model (i.e. coupled system) as
\begin{equation}
\delta_H(\bm{x},u)=\delta_L(M_P) + E(\bm{x},u)
\end{equation}
where $\bm{x}=\left\{M_P,T/W,p_{cc},\alpha_P,D\right\}$ is the vector of design variables, $u=\eta$ is a realization of the aleatory random variable $U$, $\delta_H(\cdot,\cdot)$ is the inert mass fraction when coupling constraints are satisfied, $\delta_L(\cdot)$ is the low-fidelity model given by equation \ref{eq:delta_L}, and $E(\cdot,\cdot)$ is the Kriging model of the discrepancy between the two models. By introducing the low-fidelity model the propulsion/structures and the aerodynamics/structures couplings are removed. In effect, the coupling constraints are incorporated into the construction of the error model $E(\cdot,\cdot)$. Removing the couplings eliminates the need for fixed point iterations and allows the sounding rocket design to be represented as a simple feed forward system. This may substantially reduce the computational cost of uncertainty propagation relative to performing fixed point iterations for every realization of aleatory uncertainty. However, the low-fidelity model may introduce significant epistemic model uncertainty, particularly when the Kriging model is constructed based on only a small set of initial data (i.e. small design of experiment). The epistemic model uncertainty results in additional uncertainty in the final altitude and GLOW.

\paragraph{Step 2: Selecting fixed conservative values for aleatory variables}
Next, the aleatory random variable $U$ is replaced with a fixed conservative value $u_{det}$. Instead of solving the RBDO problem in \cref{eq:rbdo}, the 5th percentile of the engine efficiency is used for the conservative value. The 5th percentile was selected because the altitude is nearly a linear function of the engine efficiency and the target probability of failure is $p_F^\star=0.05$. The engine efficiency is modeled as uniformly distributed between 0.92 and 0.98 which results in a conservative value for engine efficiency of $\bm{u}_{cons}=0.923$.

\begin{figure*}
\centering
\includegraphics[height=3.25in]{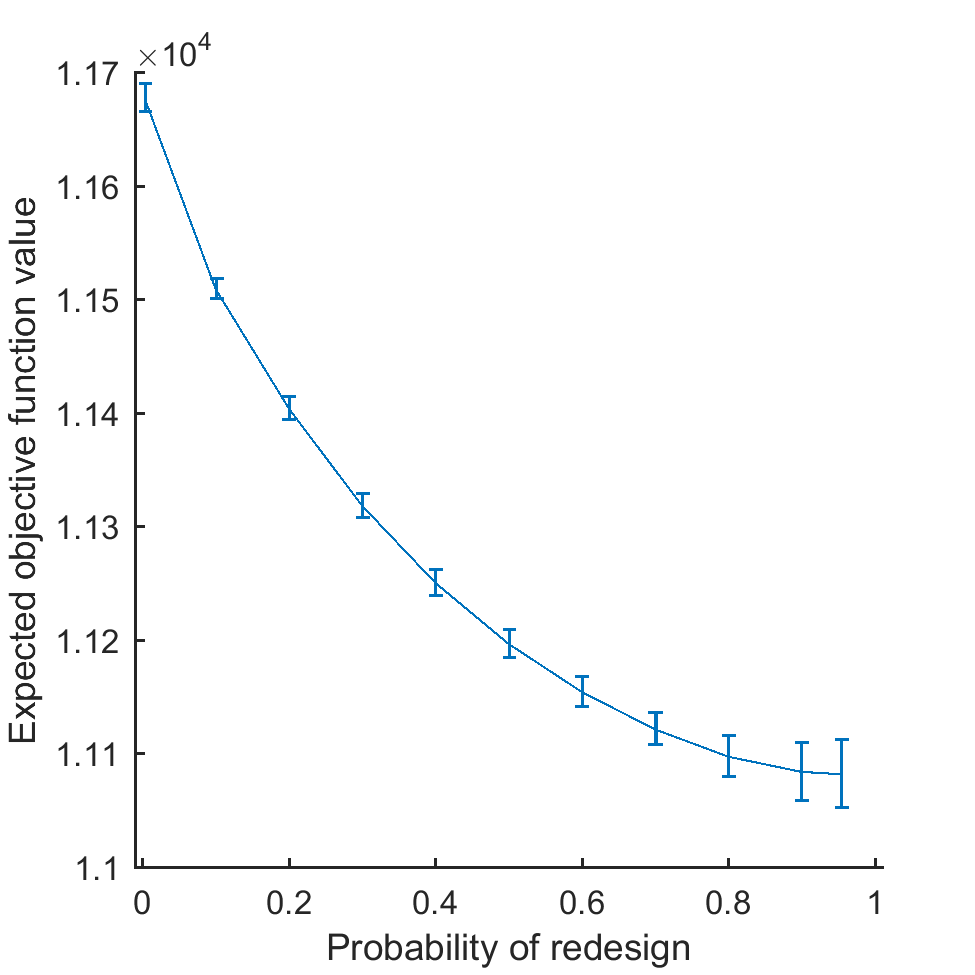}
\caption{Tradeoff curve for expected GLOW versus probability of redesign}
\label{fig:rocket_tradeoff_curves}
\end{figure*}

\paragraph{Step 3: Optimization of safety margins (i.e. standard deviation offsets)}
In the third step, the optimum standard deviation offsets are found by solving \cref{eq:saf_opt} using CMA-ES with a penalized objective function. Inside the MCS, the design optimization (\cref{eq:ddo,eq:ddo_re}) is performed using sequential quadratic programming (SQP). By varying the constraint on the probability of redesign $p_{re}^\star$ we obtain a curve for the expected GLOW versus probability of redesign as shown in \cref{fig:rocket_tradeoff_curves}. The tradeoff curve is used to determine how much risk of redesign is acceptable given the expected performance improvement. For illustration, we will select the optimum safety margins $\bm{k}=\{0.78,0.96,1.87,2.29\}$ corresponding to 20\% probability of redesign for more detailed study.

\paragraph{Step 4: Full two-level mixed uncertainty propagation}
In the fourth step, the full two-level mixed uncertainty propagation is performed for the selected optimum safety margins. The full two-level mixed uncertainty propagation is used to recover the probability of failure distribution and obtain detailed results for the MCS of the design/redesign process. For each realization of epistemic model uncertainty (i.e. Kriging conditional simulation) the probability of failure is calculated using first order reliability method (FORM). 

\paragraph{Step 5: Post-processing of simulation results}
Finally, post-processing is performed for the data gathered in the MCS. 

\begin{figure*}
\begin{subfigure}{3.25in}
  \centering
  \includegraphics[width=1\textwidth]{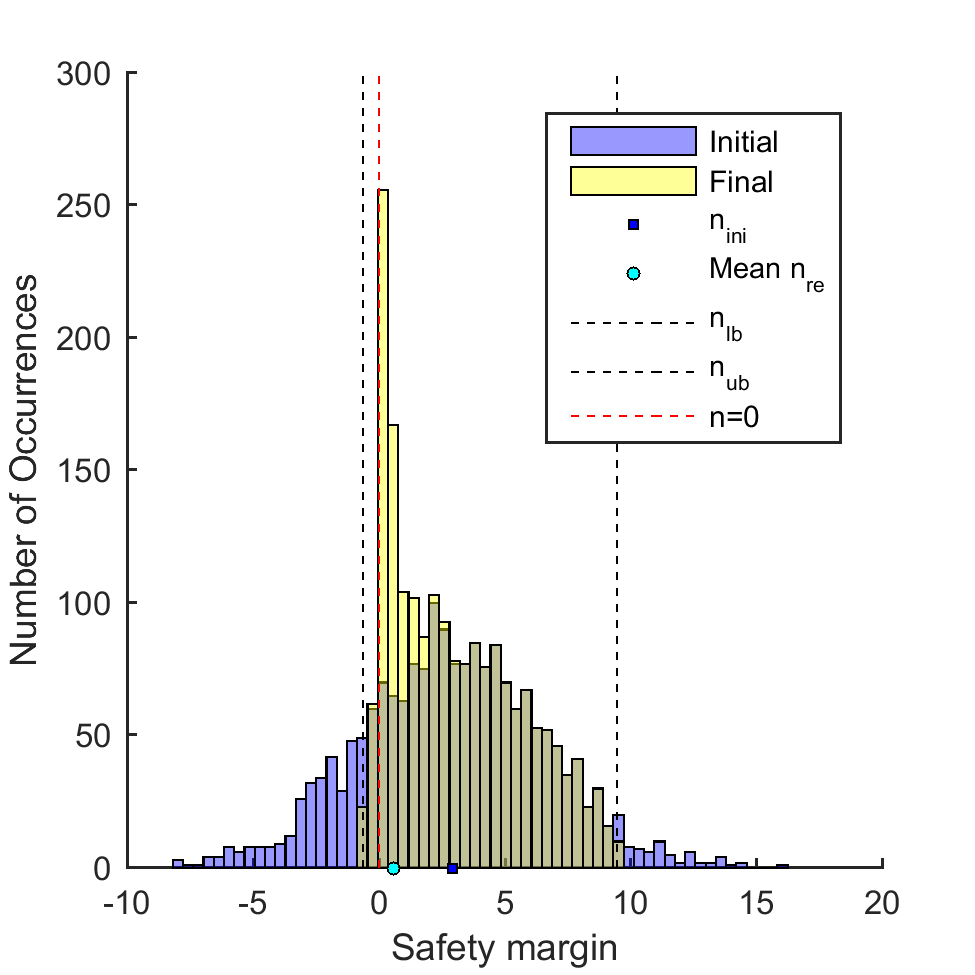}
  \caption{Safety margin}
  \label{fig:rocket_margin}
\end{subfigure}%
\begin{subfigure}{3.25in}
  \centering
  \includegraphics[width=1\textwidth]{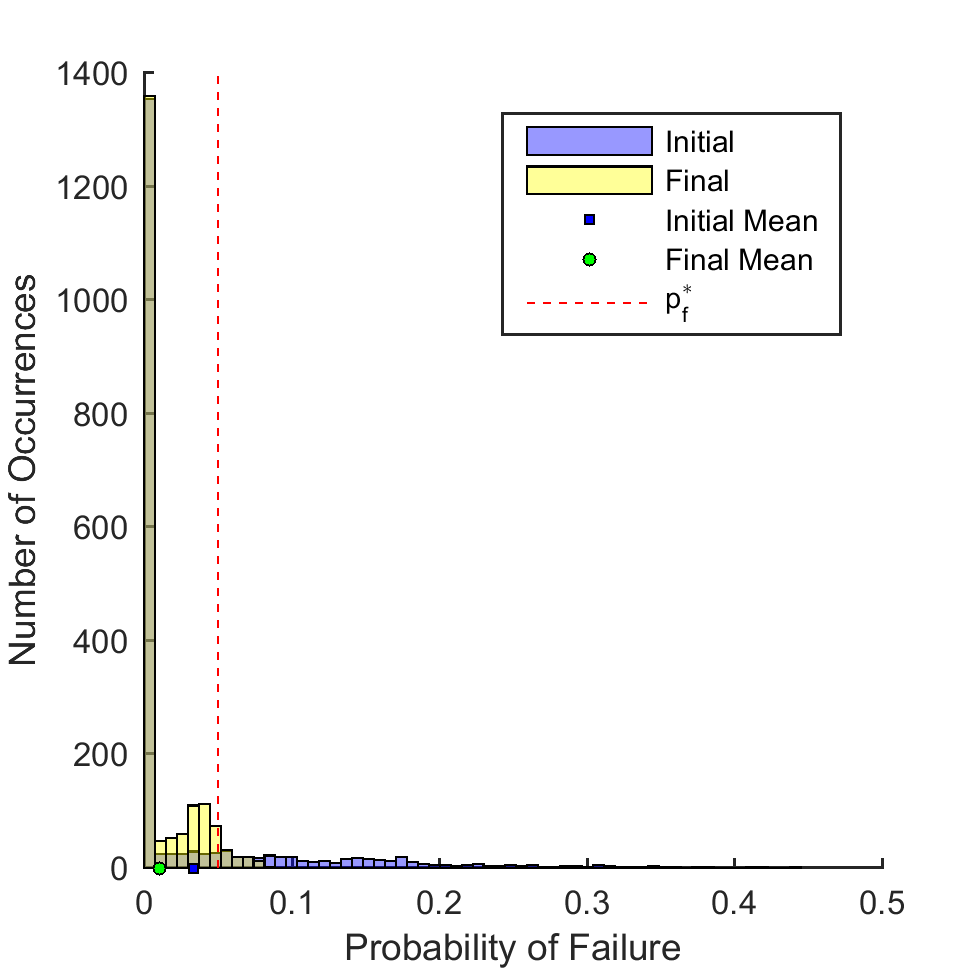}
  \caption{Probability of failure}
  \label{fig:rocket_pf}
\end{subfigure}
\caption{Distributions of safety margin and probability of failure for 20\% probability of redesign. Plots show overlapping transparent histograms.}
\label{fig:rocket_margin_pf}
\end{figure*}

First, we examine the safety margin distribution and the probability of failure distribution shown in \cref{fig:rocket_margin_pf}. The safety margin distribution in \cref{fig:rocket_margin} shows the possible constraint violations with respect to epistemic model uncertainty conditional on the fixed conservative values $\bm{u}_{cons}$. The rocket will be redesigned if the safety margin is less than $-0.6$ kilometers or greater than $9.5$ kilometers (relative to target of 150 km assuming conservative engine efficiency). Redesign acts as a type of quality control measure by initiating design changes in response to observing an extreme safety margin. We can compare the safety margin distribution in \cref{fig:rocket_margin} to the probability of failure distribution in \cref{fig:rocket_pf}. There is a strong correlation between the observed safety margin and the probability of failure (correlation coefficient -0.65). As a result, the safety margin based redesign criteria is very useful for identifying overly conservative or unsafe designs. The correlation coefficient is not as strong as in the beam example because the aleatory uncertainty in the in the engine efficiency is bounded. Due to the bounded aleatory uncertainty the correlation between safety margin and probability of failure breaks down when the safety margin is less than the point corresponding to 100\% probability of failure or the safety margin is greater than the point corresponding to 0\% probability of failure. The standard deviation offsets have been optimized based on the computationally cheap approximation of the reliability constraint in \cref{eq:pf_approx} such that the probability of a negative safety margin after possible redesign is 5\%. After performing the full two-level mixed uncertainty propagation, the probability of the probability of failure exceeding the target value of $p_F^\star=0.05$ is found to be in agreement with the target value of $\alpha=0.05$. 

\begin{figure*}[]
	\begin{subfigure}{0.3\textwidth}
    \includegraphics[width=\textwidth]{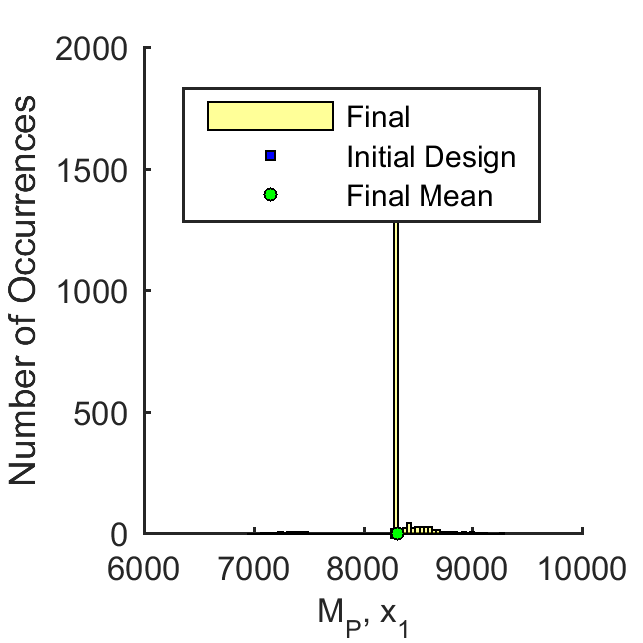}
    \end{subfigure}
	\begin{subfigure}{0.3\textwidth}
    \includegraphics[width=\textwidth]{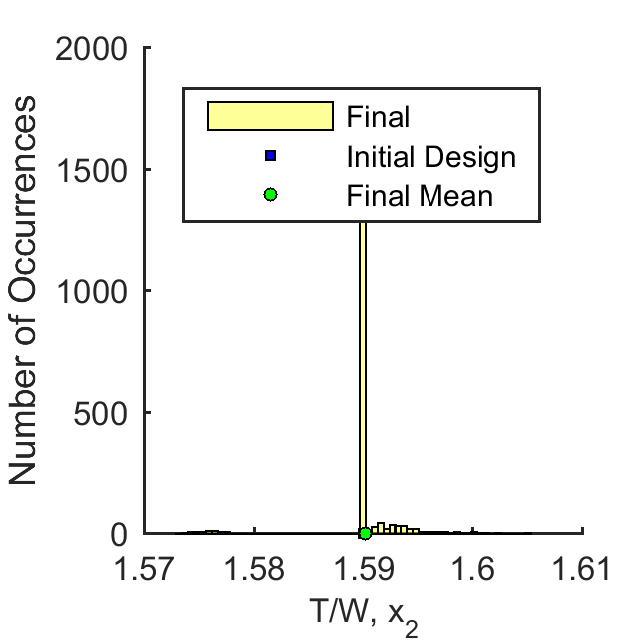}
    \end{subfigure}
	\begin{subfigure}{0.3\textwidth}
    \includegraphics[width=\textwidth]{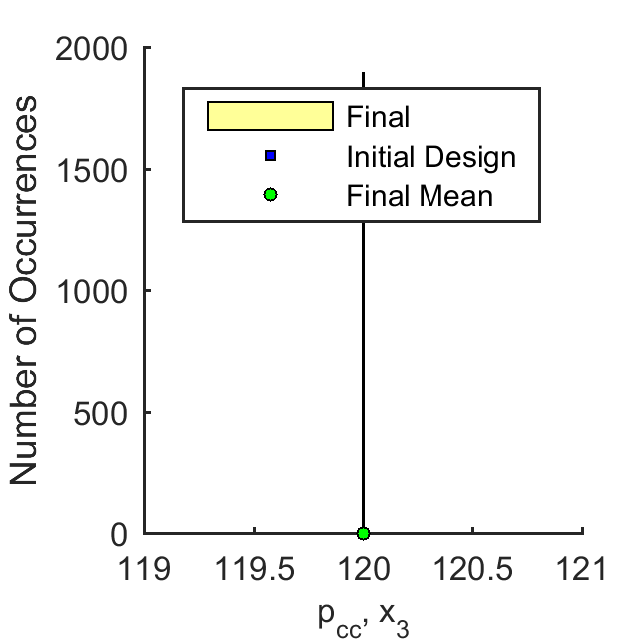}
    \end{subfigure}
	\begin{subfigure}{0.3\textwidth}
    \includegraphics[width=\textwidth]{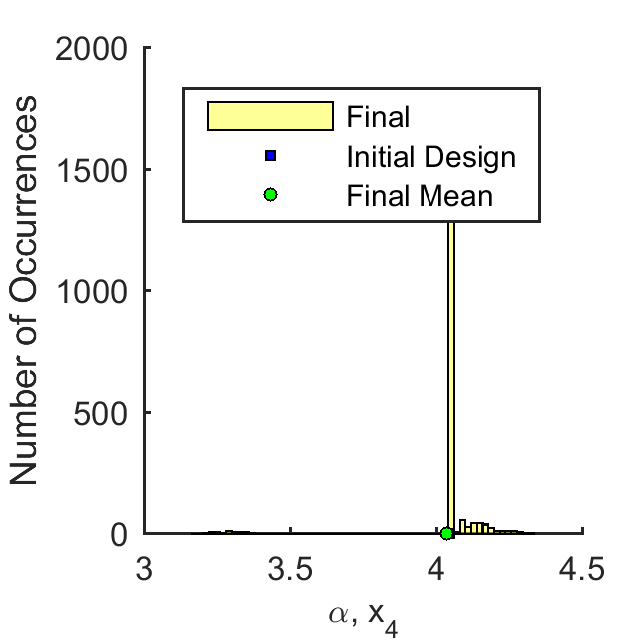}
    \end{subfigure}
	\begin{subfigure}{0.3\textwidth}
    \includegraphics[width=\textwidth]{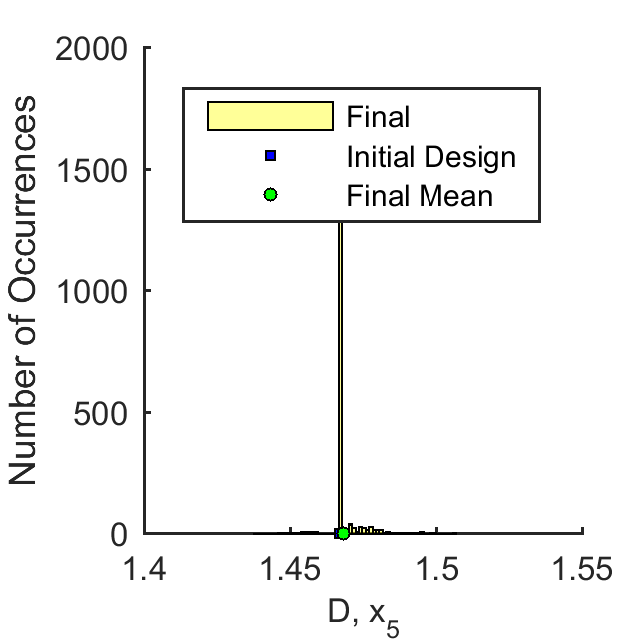}
    \end{subfigure}
\caption{Distribution of optimum design variables for 20\% probability of redesign. Plots show marginal distributions of 5-dimensional joint distribution.}
\label{fig:rocket_design}
\end{figure*}

\begin{figure*}
\begin{subfigure}{3.25in}
  \centering
  \includegraphics[width=3.25in]{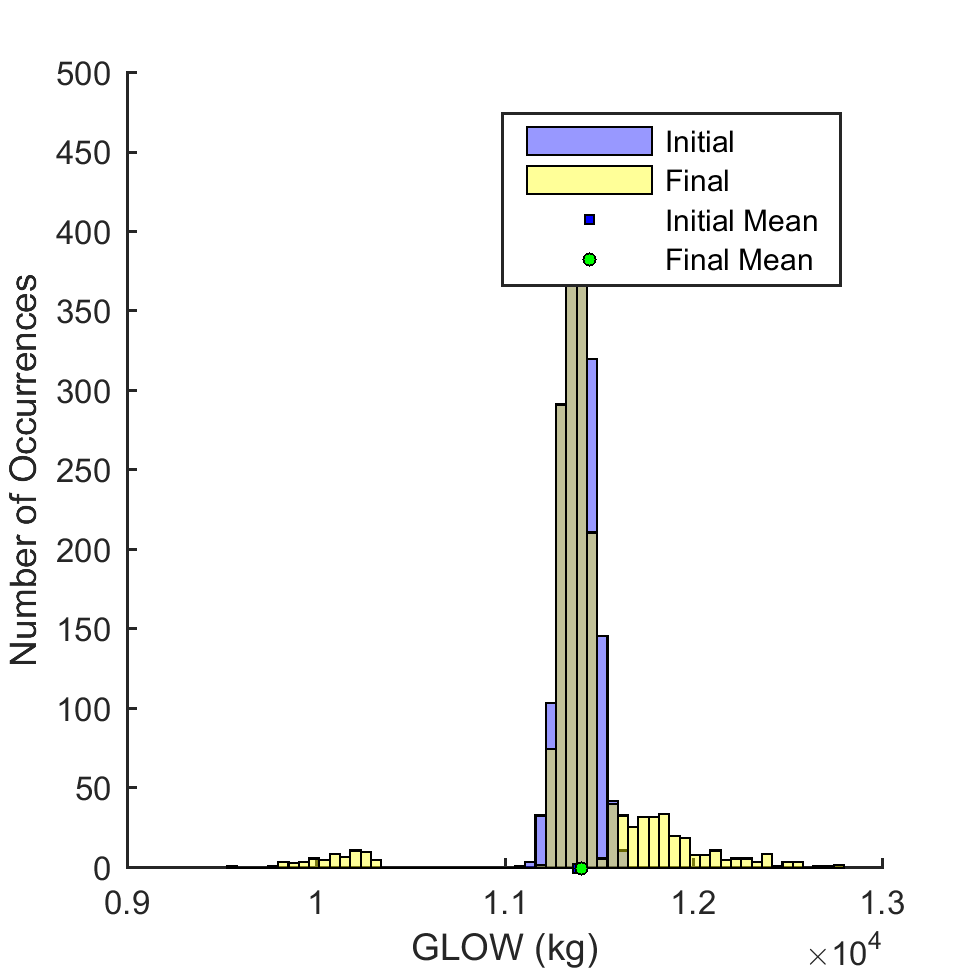}
  \caption{GLOW}
  \label{fig:rocket_glow}
\end{subfigure}%
\begin{subfigure}{3.25in}
  \centering
  \includegraphics[width=3.25in]{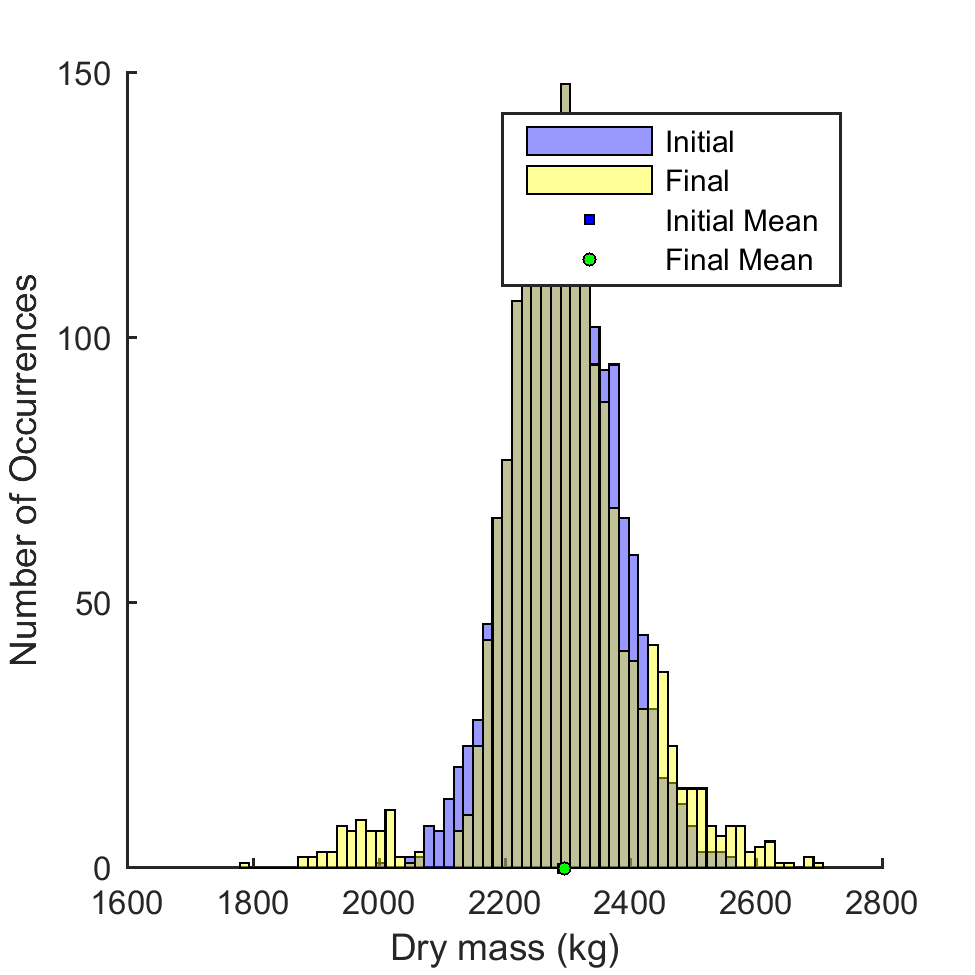}
  \caption{Dry mass}
  \label{fig:rocket_dry}
\end{subfigure}
\caption{Distributions of GLOW and dry mass for 20\% probability of redesign. Plots show overlapping transparent histograms.}
\label{fig:rocket_glow_dry}
\end{figure*}

Second, we examine the optimum design variable distribution shown in \cref{fig:rocket_design} and the GLOW and dry mass distributions shown in \cref{fig:rocket_glow_dry}. The design variable distribution is 5-dimensional so the marginal distributions are shown. The peak corresponds to the initial design since there is an 80\% probability the initial design will be accepted as the final design. The distribution of design variables is useful for planning for future design changes. It is observed that the chamber pressure does not change during redesign. The optimum chamber pressure is always the upper bound of 120 bars regardless of the outcome of the future high-fidelity evaluation. The change in diameter is relatively small with a change on the order of $\pm$1\% if redesign is required. However, the propellant mass may change substantially. The mass of propellant may decrease approximately 12\% if redesign for performance is required or increase by 4\% if redesign for safety is required. The relative change in GLOW due to redesign is similar to the relative change in propellant mass as seen in \cref{fig:rocket_glow}. The dry mass distribution is shown in \cref{fig:rocket_dry}. If redesign for safety is required, the dry mass will increase by about 2\%. If redesign for performance is required, the dry mass will decrease by about 7\%. Since underestimating the mass corresponds to overestimating the altitude, and vice versa, redesign tends to increase the mass of heavier mass realizations or decrease the mass of lighter mass realizations by adjusting the propellant mass accordingly.

\section{Discussion \& Conclusions}
\label{sec:conclusion}
At the initial design stage, engineers often must rely on low-fidelity models with high epistemic model uncertainty. One approach to high epistemic model uncertainty is to add a safety margin, such as a $k$ standard deviation offset, to design constraints to ensure the optimum design is well within the safe design space. If the safety margin is large then the designer has more confidence that the design is safe, but design performance suffers. If the safety margin is small then the design space is larger and designs with better performance become accessible, but the designer has less confidence in the safety of the design. If there will be an opportunity in the future to evaluate the design using higher fidelity modeling (or to perform a test on a prototype), then this provides an opportunity to redesign (i.e. correct or modify) a design that is revealed to be too conservative or unsafe. 

In this study we propose a safety-margin-based method for design under mixed epistemic model uncertainty and aleatory parameter uncertainty. The method is based on a two stage design process where an initial design is selected based on low-fidelity modeling, but there will be an opportunity in the future to evaluate the design with a high-fidelity model and if necessary calibrate the low-fidelity model and perform redesign. The design optimization is performed deterministically based on fixing the aleatory variables at the MPP of the mean low-fidelity model and applying a $k$ standard deviation offset to constraint functions to compensate for model uncertainty. A MCS is performed with respect to epistemic model uncertainty based on conditional simulations of a Kriging model. By repeating the determinstic design process for many different realization of model uncertainty it is possible to predict how future redesign may change the design performance and reliability. It is shown that future redesign acts similar to quality control measures in truncating extreme values of epistemic model uncertainty. The simulation allows the designer to tradeoff between the expected design performance and the risk of future redesign while still achieving a specified confidence level in the reliability of the final design. It is found that redesign for safety is particularly effective at truncating high probabilities of failure and therefore allows for improved design performance of the initial design by being less conservative. On the other hand, redesign for performance allows a designer to improve the performance of the initial design if it is later revealed to be too conservative. It is found that the optimum design strategy includes some probability of both redesign for safety and redesign for performance.

The method is demonstrated on a cantilever beam bending example and then on a multidisciplinary sounding rocket design example. In both examples it is shown that there is a strong correlation between the safety margin and the probability of failure. Therefore, the simple safety margin based redesign criteria is useful for identifying an unsafe or overly conservative design. This type of quality control measure is already incorporated into many engineering design applications. The proposed method allows for more detailed study of the effects of redesign and allows the designer to plan for future design changes and explore the interactions between the probability of redesign, safety margins, design performance, and probability of failure.

\section*{Appendix}

\section*{Acknowledgments}
Funding for this research was provided by ONERA - The French Aerospace Lab. Thanks to Dr. Brevault from ONERA for the launch vehicle problem definition used in this study.

\bibliography{myBib}
\bibliographystyle{aiaa}

\end{document}